\documentclass[]{spie}  

\newcommand{\asp}{\textit{Aspera}}
\usepackage{amsmath,amsfonts,amssymb}
\usepackage{graphicx}
\usepackage{booktabs}

\usepackage[colorlinks=true, allcolors=blue]{hyperref}

\newcommand{\OVI}{\hbox{{\rm O}\kern 0.1em{\sc vi}}}

\title{Optical alignment of contamination-sensitive Far-Ultraviolet spectrographs for Aspera SmallSat mission}
\author[a,c]{Aafaque R. Khan}
\author[a]{Erika Hamden}
\author[a]{Haeun Chung}
\author[c]{Heejoo Choi}
\author[a,c]{Daewook Kim}
\author[a]{Nicole Melso}
\author[d]{Keri Hoadley}
\author[a]{Carlos J. Vargas}
\author[b]{Daniel Truong}
\author[b]{Elijah Garcia}
\author[a]{Bill Verts}
\author[a]{Fernando Coronado}
\author[a]{Jamison Noenickx}
\author[b]{Jason Corliss}
\author[a]{Hannah Tanquary}
\author[a]{Tom Mcmahon}
\author[b]{Dave Hamara}
\author[a]{Simran Agarwal}
\author[e]{Ramona Augustin}
\author[a]{Peter Behroozi}
\author[a]{Harrison Bradley}
\author[c]{Trenton Brendel}
\author[f]{Joe Burchett}
\author[a]{Jasmine Martinez Castillo}
\author[m]{Jacob Chambers}
\author[g]{Lauren Corlies}
\author[d]{Greyson Davis}
\author[h]{Ralf-J{\"u}rgen Dettmar}
\author[a]{Ewan Douglas}
\author[i]{Giulia Ghidoli}
\author[a]{Alfred Goodwin}
\author[b]{Walter Harris}
\author[i]{Carl Hergenrother}
\author[j]{J. Christopher Howk}
\author[a]{Miriam Keppler}
\author[a]{Nazende Ipek Kerkeser}
\author[i]{John N. Kidd Jr.}
\author[a]{Jessica S. Li}
\author[a]{Gabe Noriega}
\author[a]{Sooseong Park}
\author[a]{Ryan Pecha}
\author[a]{Cork Sauve}
\author[k]{David Schiminovich}
\author[i]{Sanford Selznick}
\author[l]{Oswald Siegmund}
\author[c]{Rebecca Su}
\author[b]{Sumedha Uppnor}
\author[a]{Jacob Vider}
\author[a]{Ellie Wolcott}
\author[b]{Naomi Yescas}
\author[a]{Dennis Zaritsky}

\affil[a]{Steward Observatory, University of Arizona, 933 N. Cherry Avenue, Tucson, AZ 85721, USA}
\affil[b]{Lunar and Planetary Laboratory, University of Arizona, 1629 E. University Blvd. Tucson, AZ 85721, USA}
\affil[c]{Wyant College of Optical Sciences, University of Arizona, 1630 E University Blvd, Tucson, AZ 85721, USA}
\affil[d]{Department of Physics and Astronomy, University of Iowa, 203 Van Allen Hall Iowa City, Iowa 52242, USA}
\affil[e]{Space Telescope Science Institute, 3700 San Martin Drive, Baltimore, MD 21218, USA}
\affil[f]{Department of Astronomy, New Mexico State University, Las Cruces, NM 88001, USA}
\affil[g]{Adler Planetarium, 1300 S DuSable Lake Shore Dr., Chicago, IL 60605, USA}
\affil[h]{Ruhr-Universit{\"a}t Bochum, Universit{\"a}tsstrasse 150, 44801 Bochum, Fakult{\"a}t f{\"u}r Physik und Astronomie, Astronomisches Institut (AIRUB), Germany}
\affil[i]{Ascending Node Technologies, LLC, 2548 E. 4th Street,Tucson, AZ 85721, USA }
\affil[j]{Department of Physics and Astronomy, University of Notre Dame, Notre Dame, IN 46556, USA}
\affil[k]{Department of Astronomy, Columbia University, 550 W. 120th Street MC 5246, New York, NY 10027, USA}
\affil[l]{Sensor Sciences, LLC, 3333 Vincent Road, Suite 103 Pleasant Hill, CA 94523, USA}
\affil[m]{Department of Computer Science,  University of Arizona, 1040 4th St, Tucson, AZ 85721, USA}

\authorinfo{Further author information: (Send correspondence to Aafaque R. Khan)\\A.R.K.: E-mail: arkhan@arizona.edu, Telephone: 1 520 223 5171}

\begin{document} 
\maketitle

\begin{abstract}
\asp \ is a NASA Astrophysics Pioneers SmallSat mission designed to study diffuse {\OVI} emission from the warm-hot phase gas in the halos of nearby galaxies. Its payload consists of two identical Rowland Circle-type long-slit spectrographs, sharing a single MicroChannel plate detector. Each spectrograph channel consists of an off-axis parabola primary mirror and a toroidal diffraction grating optimized for the 1013-1057 {\AA}
bandpass. Despite the simple configuration, the optical alignment/integration process for Aspera is challenging due to tight optical alignment tolerances, driven by the compact form factor, and the contamination sensitivity of the Far-Ultraviolet optics and detectors. In this paper, we discuss implementing a novel multi-phase approach to meet these requirements using state-of-the-art optical metrology tools. For coarsely positioning the optics we use a blue-laser 3D scanner while the fine alignment is done with a Zygo interferometer and a custom computer-generated hologram. The detector focus requires iterative in-vacuum alignment using a Vacuum UV collimator. The alignment is done in a controlled cleanroom facility at the University of Arizona.
\end{abstract}


    

\section{INTRODUCTION}
\label{sec:intro} 

The Circumgalatic Medium (CGM) is a multi-phase, multi-scale gas reservoir around galaxies that roughly extends from the outskirts of the stellar disk to the viral radii of the dark matter halo \cite{Tumlinson_2017}. It is the interface where the gas accreted from the intergalactic medium (IGM) enters the galactic halo, where metal-rich gas ejected by young stars, supernovae, and active galactic nuclei (AGNs) intermixes with the gas at virial temperature, and where the gas eventually cools and falls into the stellar disk to feed the formation of stars \cite{Keres_2005,Fielding_2017,Putman2017}. This exchange and regulation of gas play an important role in the evolution and composition of galaxies. Understanding the mass budget, morphology, and dynamics of the CGM is a crucial step in advancing our knowledge of galaxy formation and evolution. 

\asp \ is a NASA Astrophysics Pioneers SmallSat Mission designed to detect and map {\OVI} emission from the warm-hot (T$\sim10^5-10^6$ K) component of CGM in halos of nearby galaxies \cite{Chung2021}. It is designed to operate in the Far-Ultraviolet region of the spectrum with a narrow spectral bandpass of $\sim$10 \AA \ and approximately centered at the rest-frame wavelength of the {\OVI} emission line at 1032 \AA. With a spectral resolution R $\sim$ 2300 \asp \ can distinguish the {\OVI} emission line from the CGM of target galaxies from the Geocoronal emission lines\cite{Feldman_2001}, and the emission from the Milky Way halo \cite{Otte_2006}. All the targets for \asp \ are high-inclination, edge-on galaxies. This selection was made to distinguish between the gas emission lines from the Halo and the stellar disk of the galaxy. The spatial resolution of $\sim$ 78.5 arcsec over the 0.5-degree slit FoV\footnote{Spatial resolution $<$ 78.5" over the 0.5-deg. FoV, while the total slit length corresponds to 1-deg.} enables measurements of the distribution of the emission below and above the galactic disk plane to map the distribution of gas. During its 9-month mission, \asp \ will build up multiple days of exposure time on selected targets to ensure spectroscopic detection of {\OVI} emission, and produce 2D morphological maps of galaxy halos, for direct measurements of physical parameters such as kinematics. 

\asp \ is currently in the System Assembly, Integration, and Testing phase (Phase D) of mission implementation. The flight model of \asp \ is aligned and integrated in a FED class 1000 clean room, while the focusing of the detector and the optical performance verification is conducted in a vacuum chamber prepared to meet the molecular cleanliness required for vacuum ultraviolet optics and detector. The integrated payload is expected to be delivered to the Space Flight Laboratory at the University of Toronto Institute of Aerospace Studies (UTIAS) for assembly in the spacecraft by early 2025. It is expected to be launched into a Sun-Synchronous, high-inclination Low Earth Orbit (LEO) in the late 2025-early 2026 time frame. 

\begin{figure}[htb]
    \centering
    \includegraphics[width=0.9\textwidth]{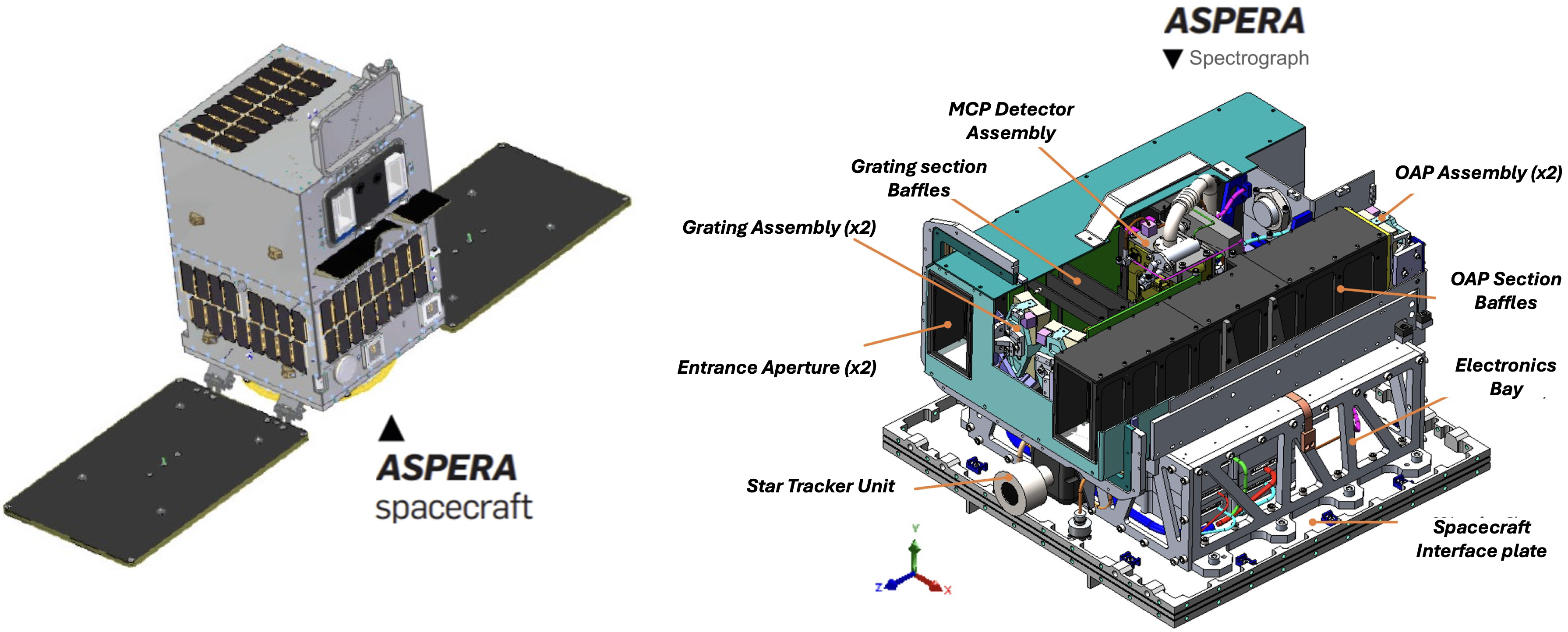}
    \caption{\asp \ Spacecraft (S/C) and Payload. (Left) \asp \ uses a customized Defiant-class SmallSat bus from the SFL at UTIAS. (Right) CAD model of the payload showing the internal components, the payload is divided into two sections the electronics bay and optical plate assembly.}
    \label{fig:spacecraft_payload_cad}
\end{figure}


This paper provides an overview of the optical design and performance of the twin spectrographs in section \ref{sec:Aspera_opical_design}. The optical performance budget and alignment tolerances (Sec. \ref{subsection:Optical Budget and tolerancing}) are also discussed. Section \ref{sec:challenges} describes the challenges and risks associated with the optical alignment of the contamination-sensitive Far-Ultravoilet (FUV) optics and the mitigations developed for \asp. Next, we provide an overview of the process (Section \ref{sec:alignment_process}) developed to address the alignment challenges and meet the instrument performance. The alignment philosophy (Sec. \ref{subsec:alignment_philosopy}) describes the top-level roadmap of the alignment process which divides it into three phases: CGH alignment of optics, alignment of slit and star tracker, and detector alignment (Secs. \ref{subsubsec: Phase_1}, \ref{subsubsec:Phase_2}, and \ref{subsubsec:Phase_3}) depending on the objective of the process and meteorology methods. The optics and slit alignment phases are conducted in a class 1000 cleanroom, while the detector alignment is done in a vacuum UV (VUV) test chamber. In section \ref{sec:Prototype_testing}, we describe the simulation and prototype testing done to validate the alignment process. The status of the flight alignment is discussed in section \ref{sec:status_futurework}. 

\section{Optical Design and Alignment Requirements}
\label{sec:Aspera_opical_design}
\subsection{Optical Design}
\label{subsection: }

\begin{figure}[htb]
    \centering
    \includegraphics[width=1\textwidth]{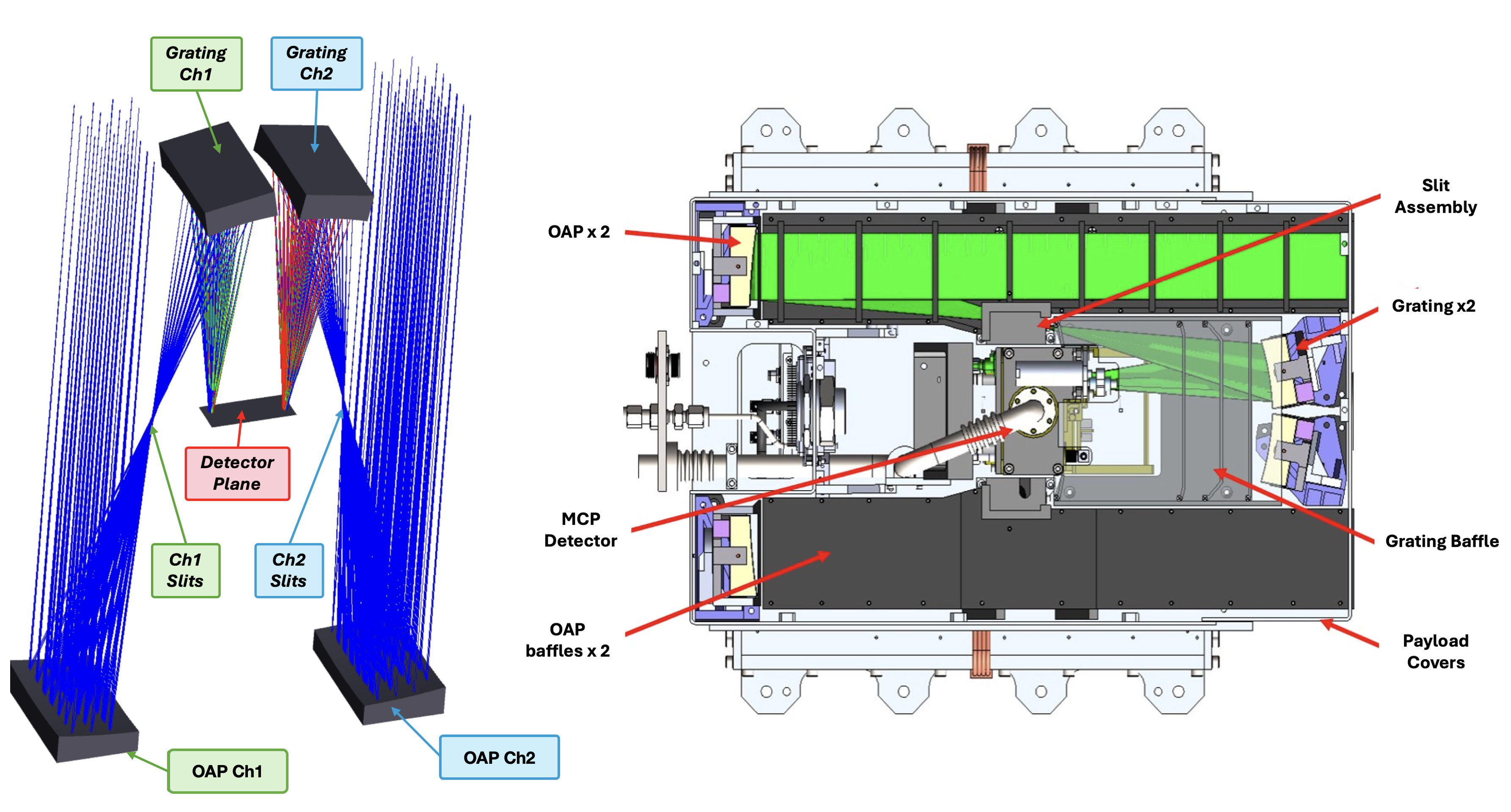}
    \caption{Optical layout and CAD model for Aspera payload. (Left) The configuration of twin Rowland circle spectrographs is mirrored to share a single Micro-channel plate detector. (Right) The two channels are offset in the spatial direction (y-axis) to avoid overlap of their spectral footprints on the shared detector. The CAD model shows the mechanical configuration of the payload and the location of the key optomechanical components.}
    \label{fig:optical_layout}
\end{figure}

{\asp} consists of two identical Rowland Circle-like spectrographs that are configured to share one Micro-channel Plate detector. The optical layout of the twin-spectrographs is shown in Figure \ref{fig:optical_layout}. The two channels have different entrance apertures but are co-aligned to have overlapping Field of View (FoV). For each channel, the beam from the sky is focused onto a long slit by an Off-Axis Parabola (OAP). The long slit acts as a spatial filter and selects a $1^{\circ}$ x 30'' FoV. The toroidal grating disperses the diverging beam from the slit and refocuses the first diffraction order on the detector focal plane. The other orders from the grating are blocked from reaching the detector by carefully designed baffles \cite{Agarwal+24}. The two channels share a Cross-Delay Line (XDL) Microchannel plate (MCP) detector with two distinct $\sim$ 3.8 mm x 3.1 mm footprints on the MCP. A detailed description of the optical design and performance modeling for Aspera is provided in Chung et al.\cite{Chung+24}. The spectrographs are designed to achieve a spectral resolution requirement of R $>$ 1500 and a spatial resolution requirement of 120" within the narrow spectral range of 1030-1040 \AA.  

\subsection{Optical Alignment Budget and Tolerances}
\label{subsection:Optical Budget and tolerancing}
\begin{figure}[ht]
    \centering
    \includegraphics[width=0.9\textwidth]{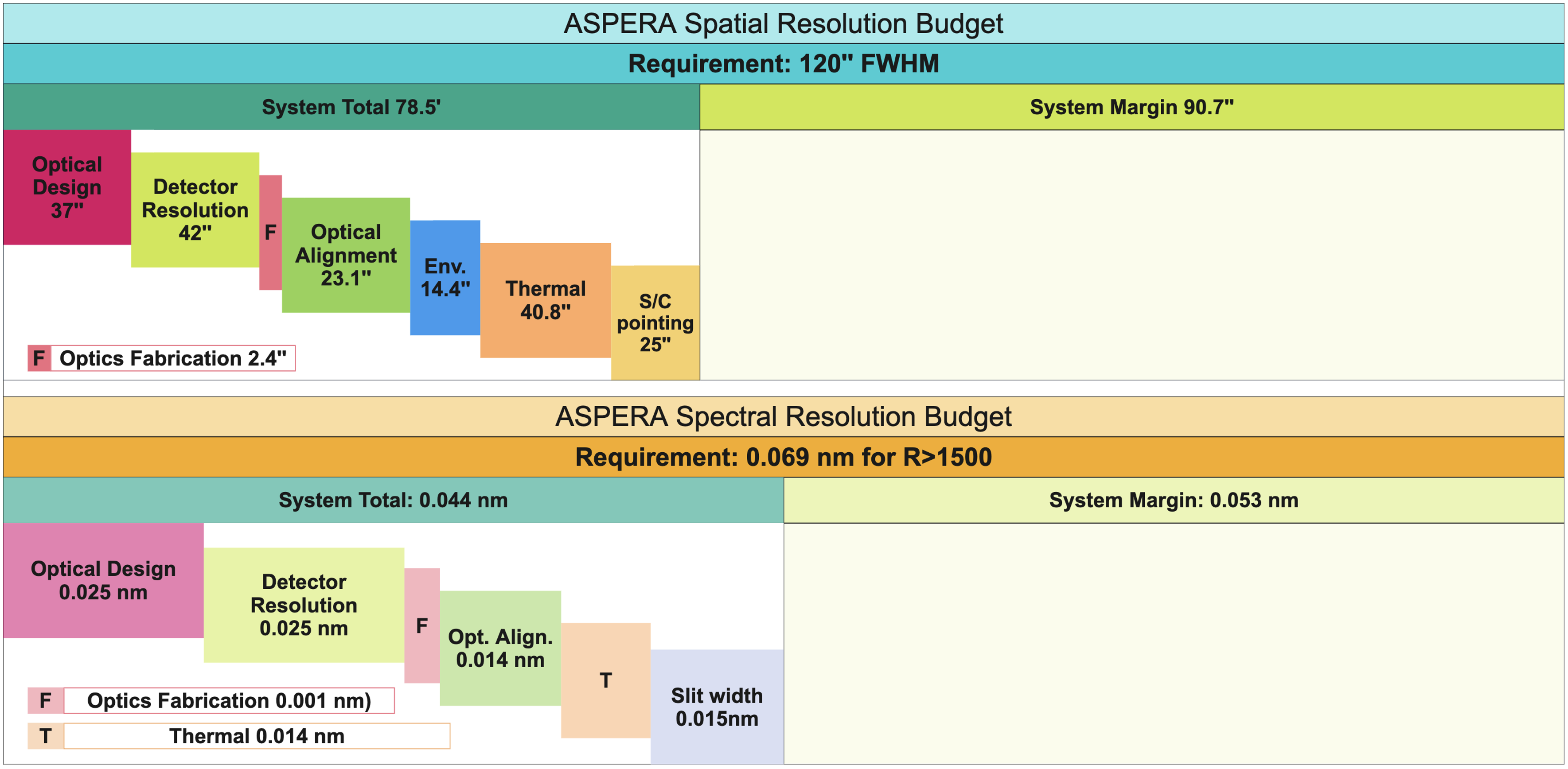}
    \caption{Spatial and Spectral resolution budget for \asp. The budget component for optical alignment was used to constrain the optical tolerances.}
    \label{fig:Optical_Budget}
\end{figure}

\begin{figure}[ht]
    \centering
    \includegraphics[width=0.9\textwidth]{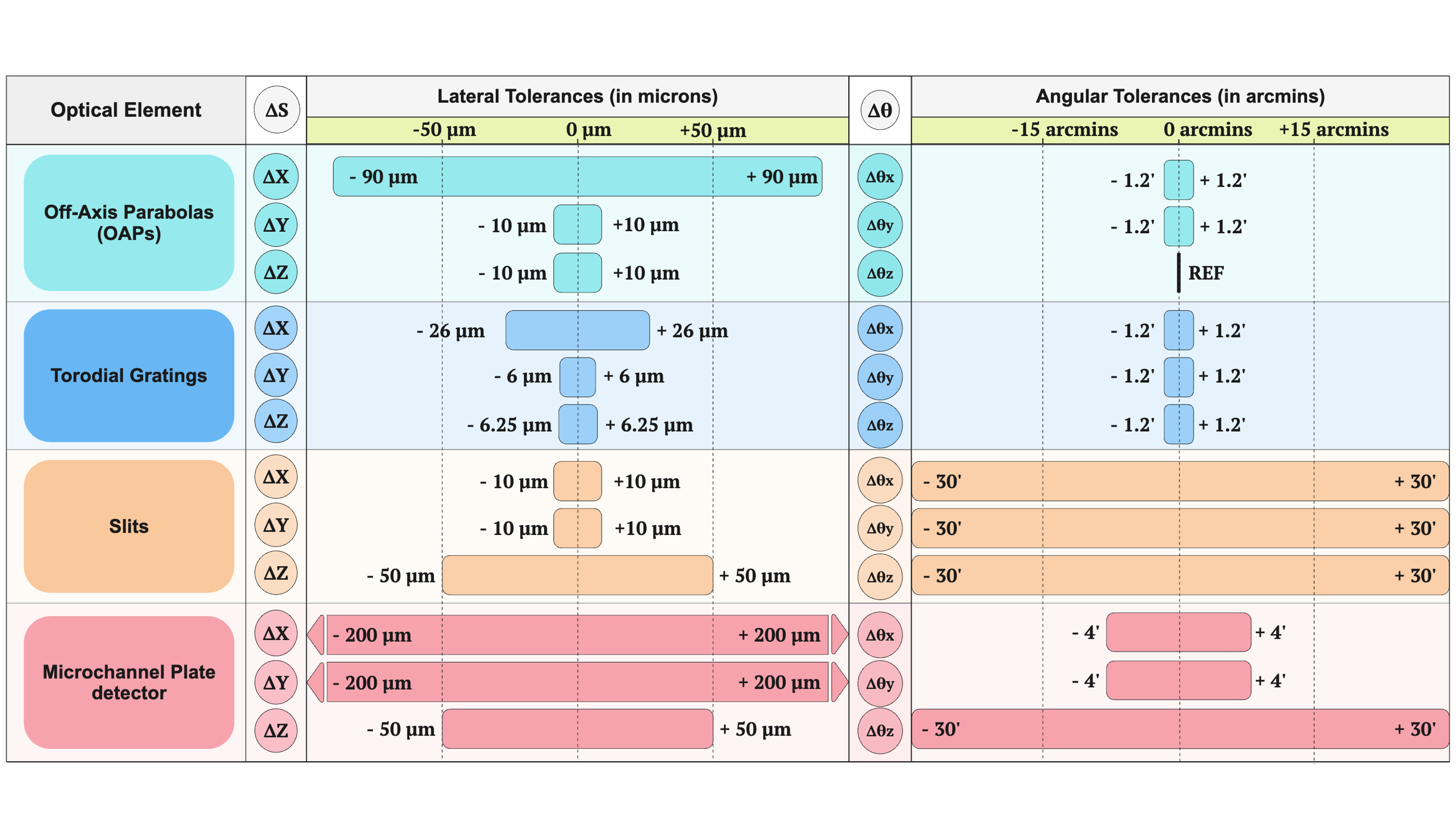}
    \caption{Summary of the alignment tolerances for the \asp \ optics derived from Monte-Carlo simulations.}
    \label{fig:Tolerance_summary}
\end{figure}

The alignment budget for Aspera is shown in Figure \ref{fig:Optical_Budget}. It gives the breakup of the expected contributions of the errors including fabrication errors, alignment tolerances, degradation due to thermal and environmental factors, and the spacecraft pointing errors to the spatial and spectral resolution performance of the instrument. The dependence of the optical performance on fabrication and alignment tolerances of the optical elements were studied using tolerance sensitive analysis and Monte Carlo simulations \cite{choi2023} in Ansys Zemax OpticStudio\textsuperscript{\textregistered}. The opto-mechanical alignment tolerances derived from these simulations are shown in Figure \ref{fig:Optical_Budget}. Despite the simple optical design, \asp \ has tight alignment tolerances that are driven by constraints on the payload volume and the co-alignment of two channels (shown in Figure \ref{fig:Tolerance_summary}). This motivates the novel alignment process that we have developed to integrate and align the spectrographs. 
\section{Optical alignment challenges}
\label{sec:challenges}
The challenges and risks that were expected during the optical alignment process of \asp \ were identified during the initial payload design. The mitigation for each challenge was developed as a part of the design process to reduce the risk profile of the mission. Table \ref{tab:challenges} provides a summary of these challenges and the mitigation solutions developed to address them. The challenges are discussed in detail below: 
\begin{itemize}
\item \textbf{Tight optical alignment tolerances:} From an optical meteorology and positioning point of view, the alignment tolerances for \asp \ optics, shown in Figure \ref{fig:Tolerance_summary}, are quite challenging. The alignment tolerances on all constrained degrees of freedom (5 for OAP, 6 for Gratings, and 3 for Detector) require high precision adjustments with the tightest specifications requiring lateral and angular step sizes of $\leq 3 \mu$m  and $\leq$ 12 arcsecond, respectively. To achieve this precision, a multi-phase, iterative process has been developed to position the optical elements and stake them in position.  

\item \textbf{Particulate and molecular contamination sensitivity of optics: } \asp's \ FUV optics are sensitive to degradation due to both molecular and particulate contamination\cite{Plesseria2001,Tribble1996}. Molecular contaminants can reduce the throughput of the instrument by absorbing the radiation at FUV wavelengths. Certain molecular contaminants can polymerize to form long-chain hydrocarbons when exposed to UV. The polymerized residues can reduce the throughput and increase the scatter of contaminated optical surfaces. Particulate contamination can obscure the optical surfaces and block the pores in the Microchannel plates in the detector leading to throughout loss and scatter. To protect the optical elements and instrument from contamination, the payload has to be integrated into a FED Class 1000  (ISO 6) cleanroom with a strict cleanliness program to mitigate any sources of particulate and molecular contamination. The vacuum systems used for alignment and testing of the payload must follow strict molecular contamination control protocols. Details of the contamination control process for \asp \ are discussed in Melso et al\cite{Melso2024}. 
\item \textbf{Optics throughput degradation due to humidity: } Conventional Al-LiF coatings are highly susceptible to degradation when exposed to humidity\cite{Wilbrandt_2014}. \asp \ optics are coated with high-FUV reflectivity ($>$ 70\% at 1030 \AA) enhanced Lithium Flouride (eLiF) coating protected by Atomic Layer Deposition (ALD) layer of a Magnesium Floride (MgF$_2$) that are less susceptible to humidity. However, long-term exposure of eLiF coated optics in high Relative Humidity (RH $>$50\%) can lead to a gradual drop in their reflectivity \cite{Fleming:17}. The duration for which \asp \ optics can be exposed to the ambient environment of the cleanroom (ranging from RH between 30\%-50\%) is limited to a few days. During most of the alignment, integration, and testing phase the optics have to be kept in a controlled nitrogen-purged environment. 
\item \textbf{Humidity sensitivity of MCP photocathode: } The \asp \ MCP has a thin opaque film of Cesium Iodide (CsI) photocathode that leads to $>$ 37\%  Quantum Efficiency (QE) at 1030 \AA for the \asp \ flight detector.  After the photocathode coatings are applied the detector must be kept in a vacuum to prevent the degradation of CsI photocathode due to hydrolysis with water vapor in the air \cite{TREMSIN2000614}. The detector is only operated in vacuum and kept under dry nitrogen purge or vacuum during storage and transportation. 
\item \textbf{Slit alignment and star tracker alignment: } The 2D spectrum from each channel will be co-added on the ground to build the exposure for each field. This requires that the FoV of the two channels overlap within ± 10 arcseconds in the spectral direction. The total exposure for each target field to reach the threshold sensitivity for detection of {\OVI} emission is achieved by multiple visits to the same point over several days to weeks. A repeatable pointing to the same field and stability of co-alignment between the channels is required to achieve this. The alignment between the star tracker and the instrument boresight determines the offset between the pointing commanded from the ground and the FoV of the instrument. The final offset within this tolerance will be measured and corrected during the onboard astrometric calibration of the instrument\cite{khan2024calibration}. However, to minimize the time spent on astrometric calibration in orbit the star tracker is aligned to the instrument boresight during the optical alignment of the payload to within $\pm$ 15 arcmin. 
\end{itemize}

\begin{table}[ht]
\caption{Summary of Optical alignment challenges for \asp \ spectrographs}
\label{tab:challenges}
\centering
\begin{tabular}{ p{35mm}  p{115mm} }
\hline
\textbf{Challenge} & \textbf{Description (D) \& Mitigation (M)}\\
\hline
\hline
\textbf{Optical tolerances} & \textbf{D:}  Achieve alignment tolerances as low as $\pm$ 6 $\mu$m and $\pm$
 1.2'.  \newline 
 \textbf{M:} Coarse positioning of optical mounts with a laser 3D scanner.\newline
\textbf{M:} High precision alignment of optics using a CGH and a laser interferometer.
\vspace{1mm}\newline 
\textbf{D:} Detector focus in vacuum within tolerance of $\pm$ 50 $\mu$m. 
\newline \textbf{M:} Focusing the detector in a vacuum with high-quality collimated input beam from a VUV monochromator.\\
\hline
\textbf{Contamination Sensitivity} & \textbf{D:} Molecular contamination can significantly degrade the instrument throughput. \newline 
\textbf{M: }Purging optics with Ultra-High Purity (UHP) Nitrogen during the CGH alignment.
\vspace{1mm}\newline
\textbf{D:} Particulate contamination can increase scatter, cover optical surfaces, and obstruct pores on MCP detector.\newline 
\textbf{M:} Alignment in better than class 1000  clean-room; Payload bagged and purged during testing and S/C Integration.\\
\hline
\textbf{Humidity Degradation of Optics Coating} & \textbf{D:} Enhanced LiF-protected aluminum (eLiF) coatings on the optics are sensitive to humidity exposure. \newline \textbf{M:} Optics are transported in a humidity-controlled environment. \newline 
\textbf{M:} Alignment will be carried out in low humidity enclosure with limited exposure to ambient humidity of the cleanroom\\
\hline
\textbf{Water vapor sensitivity of MCP Photocathode } & \textbf{D: } CsI Photo-cathode on Micro-channel plate detector reacts with moisture leading to throughput loss. \newline
\textbf{M:} Detector stored under vacuum or Dry Nitrogen purge and operated in vacuum during alignment. \\
\hline
\textbf{Star Tracker co-alignment} & \textbf{D: }Co-alignment of the two channels within $\pm$ 10 arcsec;\newline Star tracker to instrument boresight alignment: $<15$ arcmin\\ 
&\textbf{M: } The two channels are co-aligned with each other using the CGH. \\
&\textbf{M: } The star tracker is mounted on the payload optical bench and is co-aligned with the instrument boresight reference alignment cube.\\
\hline
\hline
\end{tabular}
\end{table}
\section{OPTICAL ALIGNMENT PROCESS}
\label{sec:alignment_process}

\subsection{Alignment philosophy}
\label{subsec:alignment_philosopy}
\begin{figure}[ht]
    \centering
    \includegraphics[width=1.0\textwidth]{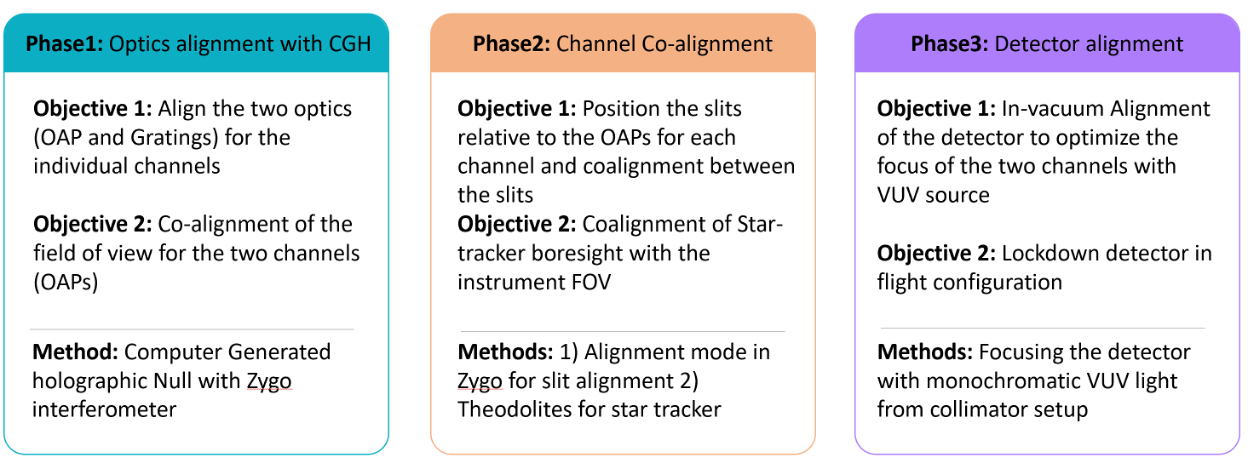}
    \caption{Optical alignment phases for Aspera: Each phase has distinct objectives and methods for achieving precision alignment. The first step of each phase is coarse positioning of the flight hardware using in-situ mounts and removable shim blocks.}
    \label{fig:alignment_phases}
\end{figure}

 A roadmap for optical alignment of \asp \ was created to address the challenges described in section \ref{sec:challenges}. As shown in Figure \ref{fig:alignment_phases}, the roadmap divides the alignment process into three phases: 1) Optics alignment with CGH, 2) Channel co-alignment, and 3) Detector alignment. The phases are distinguished by the alignment objectives and the method by which the alignment is achieved. In each phase, the tight alignment tolerances are achieved by an iterative two-step process, the first step is coarse drop-in and alignment of optomechanical assemblies using in-situ meteorology with a laser 3D scanner (Sec. \ref{subsub:3d_scanner_meterology}), and the second step is precision adjustment to align the assemblies using optical feedback. The method to get quantitative optical feedback is different for each phase. In phase 1, the optics are aligned using a Zygo Verifire\texttrademark \ Laser Interferometer with Computer Generated Hologram (CGH) nulls (Sec. \ref{subsubsec: Phase_1}). In phase 2, an imaging setup is used to place the slits at the focus of the OAP using a laser from Zygo collimated input source (Sec. \ref{subsubsec:Phase_2}). In phase 3, a collimated FUV input beam is used to focus the detector in the payload (Sec. \ref{subsubsec:Phase_3}). For precision alignment steps, the flight mounts have been designed to be adjustable with small step sizes within a limited range. The mounts for the optics have non-flight Ground Support Components (GSCs) with precision adjusters that allow rigid body movement in all six Degrees of Freedom (DoF). The detector mount has adjustability in three DoF, allowing precision movement in the defocus, tip, and tilt of the detector plane. The coarse alignment step positions the optical elements within the travel range of the flight mounts with the GSCs. The GSCs are removed after the optics are aligned and bonded in place. 

\subsection{Alignment process workflow}
\label{subsec:alignment_process}
\subsubsection{Coarse positioning and meteorology with laser 3D scanner}
\label{subsub:3d_scanner_meterology}
In the coarse positioning step of each phase, the opto-mechanical mounts are positioned by comparing the drop-in location of the assembly with the nominal location in the CAD model. For each mount, the as-built payload location is measured using a portable handheld laser 3D scanner, HandySCAN 700\textsuperscript{\texttrademark} series from Creaform, and compared directly with the CAD model using the VXinspect model in the VXelements software suite. The CAD-to-scan comparison is done using an inspection program that consists of reference measurements that are used to determine the position of optical elements to fixed datum features in the CAD model. A 3D color map showing the deviation between the CAD model and the scan of the as-built assembly is also generated by the software for comparison. This workflow is typically used in the part inspection and reverse engineering but we have adopted it for meteorology and alignment of optical components. The direct scan to CAD comparison allows coarse positioning of optomechanical mounts to within $\pm$ 75 microns (lateral) and $\pm$ 10' (angular) of the nominal location on the optical bench. 

\begin{figure}[ht]
    \centering
    \includegraphics[width=0.85\textwidth]{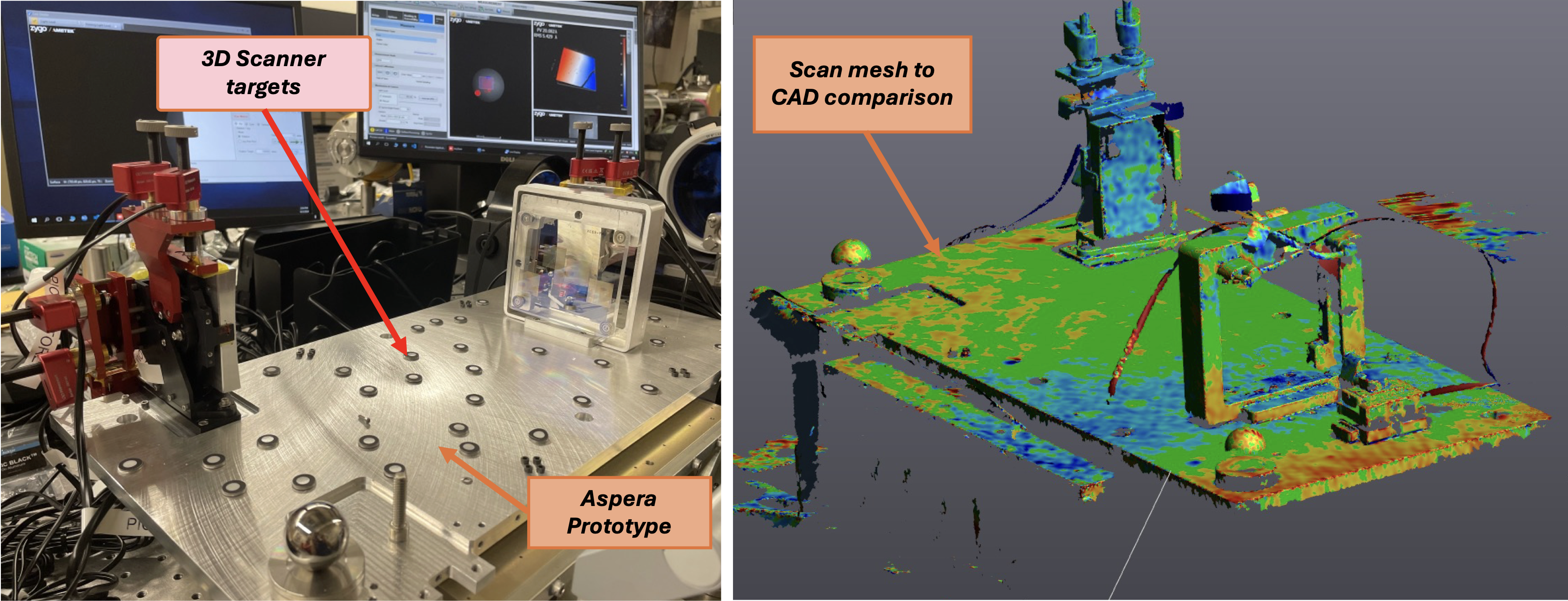}
    \caption{Example of coarse alignment positioning with 3D-scanner. (Left) \asp \ prototype test setup with CGH mount and prototype OAP mount. The alignment targets for the 3D scanner are bonded to blank washers and placed on the prototype optical plate. (Right) Example of 3D scan data from scans of the prototype bench. The color scale in this image is a comparison between two consecutive scans.}
    \label{fig:coarse_alignment}
\end{figure}

The scanner can take up to 1,300,000 measurements per second with an accuracy of up to 25 $\mu$m with volumetric accuracy 20 $\mu$m $\pm$ 40 $\mu$m/m \cite{creaform_handyscan_2024}. The scan-to-mesh conversion and inspection for each set of measurements take only a few minutes. A major advantage of this method is that it provides fast, contactless, high-precision meteorology that is crucial for minimizing the exposure of contamination-sensitive optics. As each scan of the payload optical assembly takes only a few minutes and the measurements are automated with the inspection program, the position of an optical element can be iteratively adjusted with shims till the coarse alignment is within the  range of the precision adjustment GSCs on the opto-mecahnical mounts. To integrate the into our alignment process, we tested the scanner in cleanroom environments with different flight and non-flight hardware. It can pick different surfaces ranging from highly reflective mirrors to parts coated with diffuse black optical coatings (e.g. Acktar Magic Black). The alignment targets that are used by the scanner for spatial localization during the scan are attached to washers that can be placed at convenient locations without any direct contact with the flight hardware. An example 3D scan measurement of \asp \ prototype optical bench setup is shown in Figure \ref{fig:coarse_alignment}. The features and details shown in the figure were captured within a few minutes.

\subsubsection{Phase 1: Optical alignment with CGH alignment }
\label{subsubsec: Phase_1}

In this phase, the optics (OAP and Grating) of the individual channels are to be aligned within the specified tolerances (See figure \ref{fig:Tolerance_summary}. At the same time, the two channels must be aligned with respect to each other to obtain co-pointing at the same fields in the sky. This requires the relative tilt between the channels (OAPs) to be measured and minimized during the assembly of the individual optical assemblies on the payload optical bench. This phase will be carried out in a FED Class 1000 (ISO 6). The cleanroom temperature is controlled at 293 $\pm$ 1 K. However, we do not have precise control over the humidity of the cleanroom. While the RH remains between 30\% over the year it can exceed 50\% during the monsoon season in Tuscon. To protect the coatings optical coatings from humidity in the cleanroom, the payload will be kept in a dry enclosure with constant dry nitrogen flow. The enclosure would be sealed and purged with an active humidity control system during non-working hours. 

\begin{figure}[htb]
    \centering
    \includegraphics[width=1.0\textwidth]{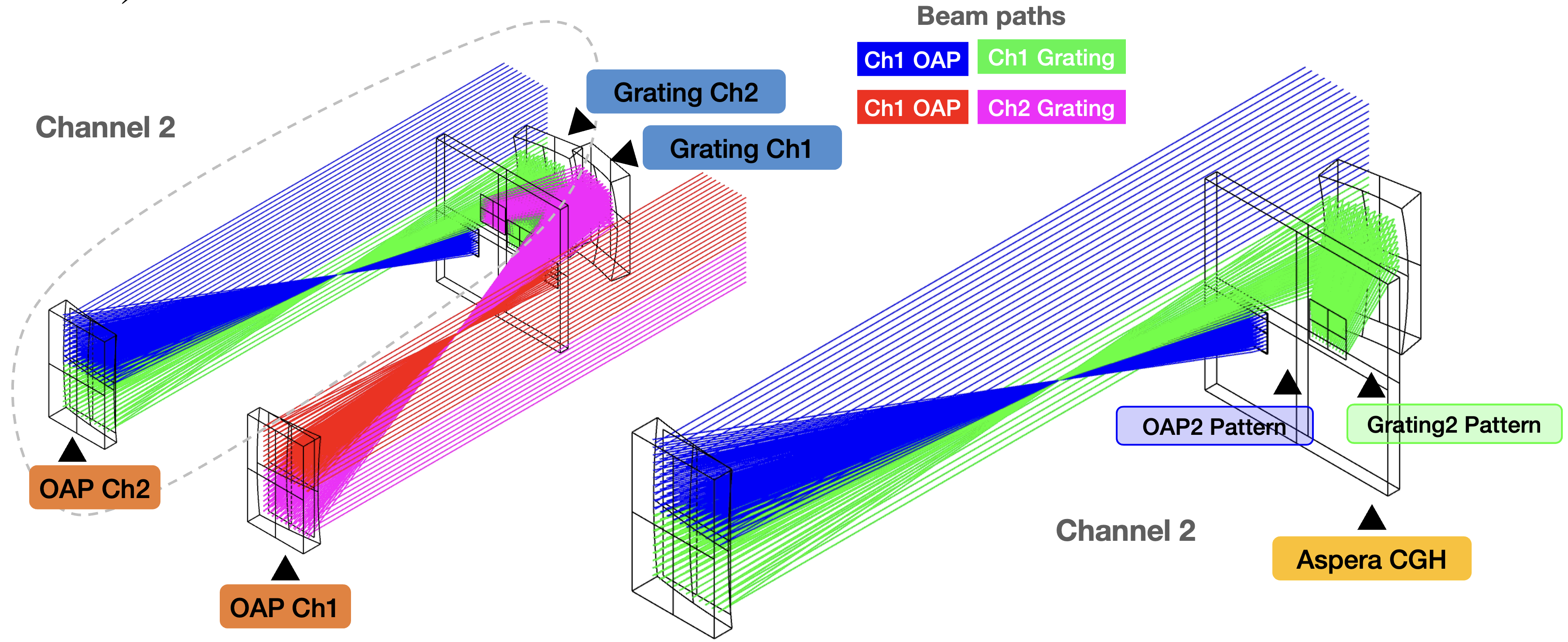}
    \caption{Layout for the CGH alignment scheme: (Left) Isometric View of the Ansys Zemax Opticstudio model showing the beam path for the CGH alignment design. For each channel, half of the collimated beam from the interferometer (Blue, Red) is returned from the OAP alignment patterns of the CGH. The other half of the beam transmits through the CGH is reflected by the grating and is returned by grating alignment patterns (Green, Pink) on the CGH. (Right) Zoomed in view of a single channel (Ch2) to show the footprints of the beams on the CGH substrate (Right).} 
    \label{fig:CGH_alignment_setup}
\end{figure}

To achieve this simultaneous alignment and metrology, we use a Zygo interferometer combined with a custom-designed CGH with four null patterns (one for each optical element). Figure \ref{fig:CGH_alignment_setup} shows the optical layout of the CGH alignment scheme designed in collaboration with the Arizona Optical Metrology (AOM) LLC. The alignment layout is designed for optical interferometry with a visible laser from Zygo and operates in the zeroth diffraction order for the \asp \ gratings. In this alignment scheme, when individual optics are placed close to their nominal design positions with respect to the CGH, a collimated beam incident into the system would be returned by the null pattern. In an ideal scenario, the optics can be positioned perfectly such that the on-axis return beam would have no wavefront error (the null condition). In the realistic case, when the optics are not perfectly aligned, the interference fringes in the off-null positions can be used to quantify the position error, in all six degrees of freedom, for the individual optics. For \asp \ flight optics the contribution of alignment errors due to rigid body displacement will dominate wavefront errors as the contribution from fabrication is expected to be small ($<$127 nm PV surface irregularity). 

\begin{figure}[htb]
    \centering
    \includegraphics[width=1.0\textwidth]{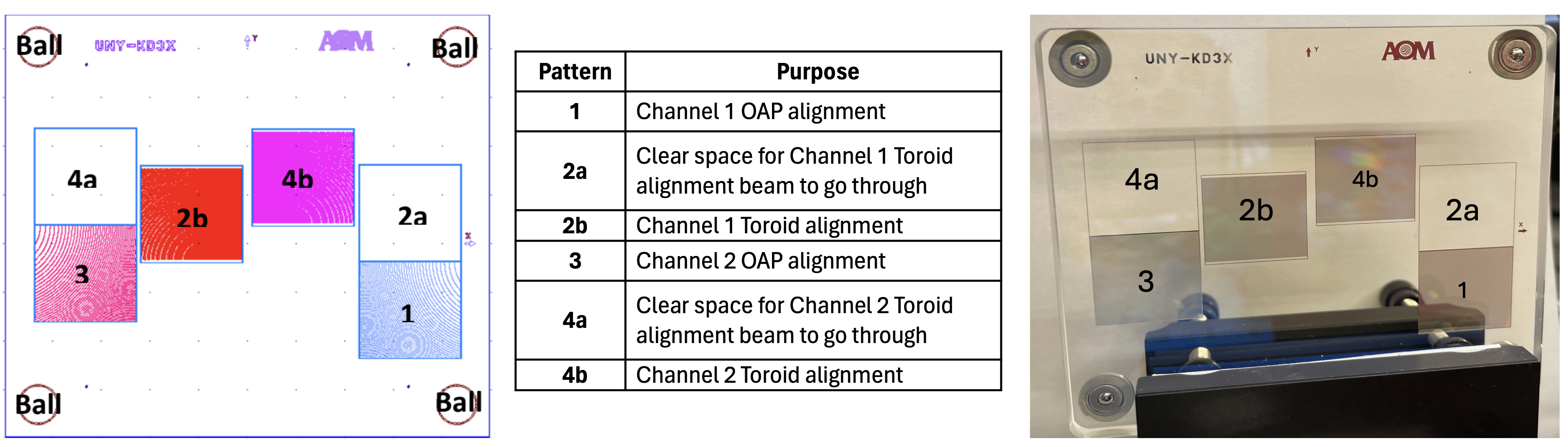}
    \caption{Aspera CGH designed and fabricated: (Left) The design of the CGH shows the layout of the null patterns for the four optics. (Right) The fused silica substrate for Aspera with the four null patterns. The four balls, that have been bonded to the substrate with high precision, are used to mount the substrate into repeatable magnetic cones. }
    \label{fig:aspera_cgh}
\end{figure}

Figure \ref{fig:aspera_cgh} shows the design of the \asp \ CGH and the fabricated CGH substrate. The four patterns are reflective amplitude CGHs that operate in the third diffraction order to avoid ghosting and artifacts in the return beam. An anti-reflection coating is applied to the substrate to further suppress ghost reflections. The wavefront error due to the pattern writing errors for the four patterns is expected to be below 10 nm RMS.

\begin{figure}[htb]
    \centering
    \includegraphics[width=0.9\textwidth]{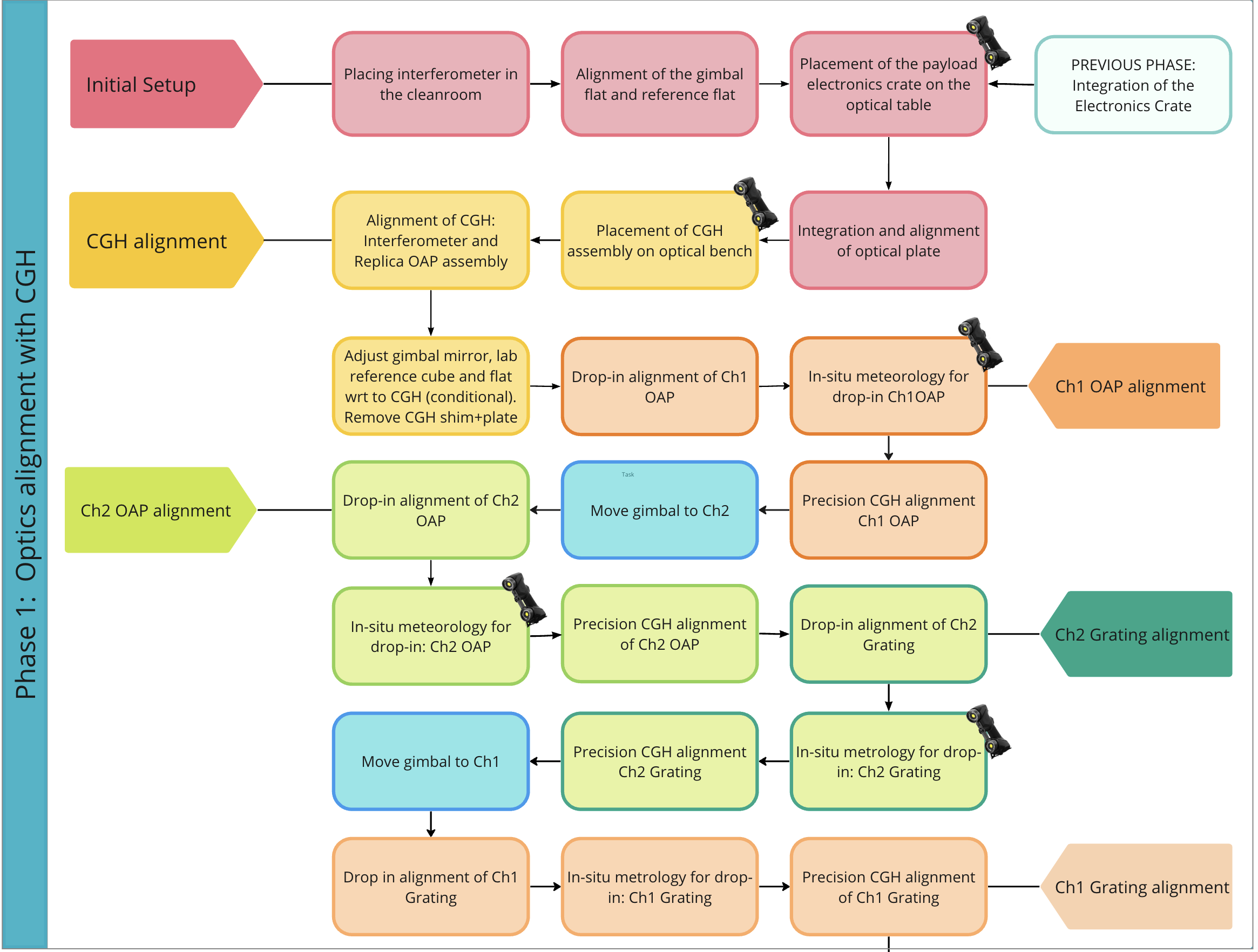}
    \caption{Phase 1 Flow Diagram: Step-by-step process for alignment of {\asp} optics with CGH. The optical assemblies are installed and aligned to the CGH sequentially. Each block in the flow diagram represents an alignment activity with a unique procedure and activity ID. The color coding of the block is to group activities that are related to specific hardware or sequence of operations. The activities with the icon with Handyscan on the top right indicate that 3D scanner measurement will be done at this stage. }
    \label{fig:phase1}
\end{figure}

The alignment flow for phase 1 is shown in Figure \ref{fig:phase1}. The CGH is aligned to the interferometer axis on a three-axis mount using the return from the substrate. The assembled mounts for the four optics are introduced one at a time and aligned to the CGH.  The individual wavefront maps for each optic are used to derive the RMS wavefront error and extract Zernike coefficients for the wavefront measured for each optical element.

Using the OptiStudio model of the setup and the Zernike coefficients derived from the wavefront measurements with Zygo, the rigid body deviations of the optical element being aligned are estimated. The flight mount for each optical assembly has six degrees of freedom with a limited travel range (approximately $\pm$150 $\mu$m in lateral and $\pm$ 30' in angular direction). The details of the flight mounts and the flight alignment results will be provided in Khan et. al. (in preparation)\cite{Khan_2025}. During alignment, the flight mounts have additional Ground support components with Piezo Linear actuator-based adjustment mechanism that allows precision movement with a step size of $<3$ $\mu$m (lateral) and $<3''$ (angular) in all six axes. These adjusters are used to move the individual optics until the RMS wavefront error and Zernike coefficients indicate that the optic is within the tolerance range shown in Fig. \ref{fig:Tolerance_summary}. At this point, the flight mounts are fixed in place by bonding with flight-compatible epoxy. The CGH will be kept in place during and post the bond cure period to measure any drift during the bonding and to verify that the post-bond position is within tolerance. After post-bond curing verification, the GSC and adjustment mechanism are removed from the flight mounts.  

\subsubsection{Phase 2: Slit and Star tracker alignment}
\label{subsubsec:Phase_2}
In this phase, the slits are installed while ensuring that the FoVs of two channels are aligned within $\pm$ 10'' in the spatial direction. This overlap between the two slits is required to co-add the data from the two slits to build Signal-to-Noise Ratio (S/N) while maintaining the spatial resolution requirement. 

\begin{figure}[htb]
    \centering
    \includegraphics[width=1\textwidth]{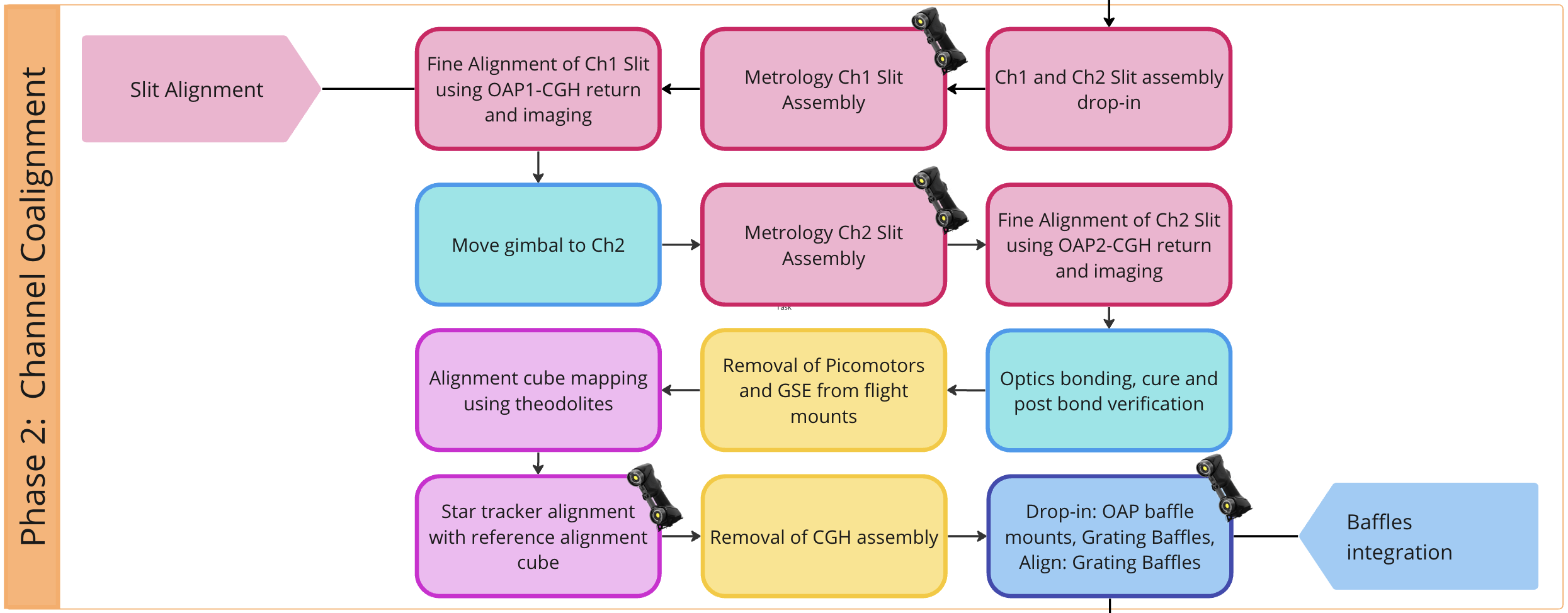}
    \caption{Phase 2 Flow Diagram: Step-by-step process for alignment for Slits and Star Tracker.  Each block in the flow diagram represents an alignment activity with a unique procedure and activity ID. The color coding of the block is to group activities that are related to specific hardware or sequence of operations. The activities with the icon with Handyscan on the top right indicates that 3D scanner measurement will be done at this stage.}
    \label{fig:phase2}
\end{figure}

The alignment flow for phase 2 is shown in Figure \ref{fig:phase2}. The slit assemblies are positioned on the optical table and the slit is coarsely positioned using the 3D scanner and shims. The slits for the two individual channels are placed in position using the interferometer return beam from the null pattern of the respective OAPs. As the optics and CGH are at fixed positions post phase 1, the interferometer beam is like the on-axis field for \asp \ with the OAP focus located at the desired center of the slit. The slit assemblies for the two channels are positioned on the optical bench using the mechanical datum features on the bench. Removable shims are used to make coarse adjustments in the drop-in position of the assembly within the clearance holes for the mount on the optical plate. The coarse alignment of the slit is achieved by adjustments when the beam passes through the slit without vignetting. The beam from OAP passing through the slit would be returned by the CGH and the return spot would be visible in the alignment mode of the interferometer. At this point, the slit would still not be in the optimal position as the OAP beam passes through with some degeneracy in clocking, de-focus, and de-centers of the slits.

\begin{figure}[htb]
    \centering
    \includegraphics[width=1\textwidth]{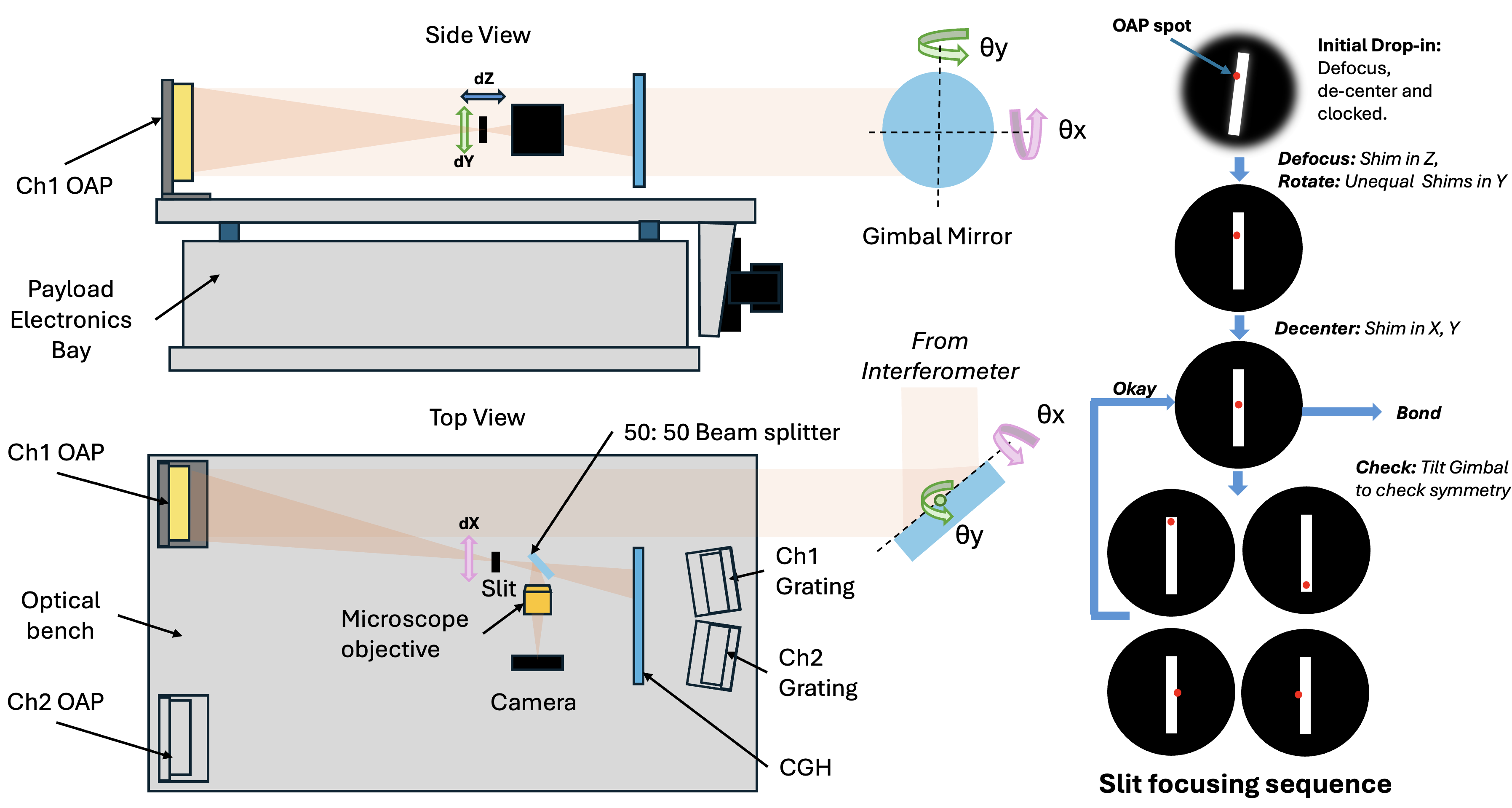}
    \caption{Schematic of the slit alignment setup and sequence: The position of the slit with respect to the OAP focus is adjusted by feedback from the imaging setup consisting of a beamsplitter, a microscope, and a compact monochromatic CMOS camera. The part of the beam transmitted through the beam splitter is returned by the CGH to the interferometer (Left). The centration of the slit in the X and Y directions is checked by moving the focused spot in the spatial and spectral directions, respectively (Right). The spot is moved to different field positions by tilting the goniometer about the Y or X axis. The CGH design allows for return from the off-axis beam which can be seen in the alignment mode of the interferometer.}
    \label{fig:slit_alignment}
\end{figure}

To get precisely place the slit at OAP focus, an imaging system is placed behind the slit such that the front focal imaging system coincides with the focal plane of the OAP. The setup consists of a pellicle beam splitter that diverts part of the beam into a microscope objective that is placed at a fixed distance from a monochromatic CMOS camera. The pellicle beamsplitter ensures that part of the beam still goes to CGH and is returned by the OAP null pattern to the interferometer. The microscope is configured such that the conjugate planes, when the microscope is placed at a predefined distance from the OAP focus, are at the OAP focal plane and the camera focal plane, respectively. Therefore, by adjusting and aligning the camera the on-axis spot from the OAP can be refocused on the microscope camera. The microscope and camera are chosen to minimize the degradation of Point Spread Function (PSF) and provide a large field of view to image the slit edges and the spot simultaneously. Figure \ref{fig:slit_alignment} shows the schematic layout of the imaging setup and a visualization of the slit alignment sequence. 

As the drop-in position of the slit is not exactly at the focal plane of the OAP, the slit edges would appear blurry in the microscope image. The slit assembly would be moved by shimming against mechanical datums in defocus direction to get the slit edges at best focus in the microscope camera. When the slit edges and spot in the camera are on the same focal plane the slit is close to the best focus spot. It is not expected that any significant tip and tilt in the slit plane can be inferred by moving the interferometer beam using the Gimbal mirror over the focal plane while tracking the spot in the microscope camera. The slit edges would get defocused depending on the direction of the tip-tilt. For precision adjustment in the X, and Y decenters and rotation, we use the fold flat between the interferometer and the payload to steer the on-axis beam within the slit FOV while still getting a retro from the OAP (and grating) CGH nulls. The edge of the slit would be detected by steering the beam and observing the vignetting in the alignment mode of the interferometer. The slit de-centers and rotation would be adjusted to make the two edges symmetric about the as-built on-axis focus of each OAP. This would be achieved iteratively, with the fold flat tip-tilt used to check the symmetry of the slit edges. Once the slit for one channel is aligned, the gimbal mirror will be moved to align the slit for the second channel. As this scheme uses the OAP retro-patterns on the CGH for the alignment of both slits, the on-sky FOV for both slits would be aligned with each other. 

After the slit installation, the star tracker will be installed. Boresight offsets between the star tracker and the instrument can be measured in orbit to correct for any alignment mismatch between the instrument's boresight spacecraft pointing. However, to allow for ease of astrometric calibration in orbit, the angle between the star tracker and instrument boresight will be aligned to be within 15 arcmin on the ground. The star tracker is mounted on the optical bench of the payload and will be installed and aligned by the payload team. 

At this point, the star tracker can be installed on its bracket under the optical bench. The star tracker has two reflective surfaces that are referenced to the center line of the star tracker boresight (see figure \ref{fig:spacecraft_payload_cad}). The two reflective surfaces will be used to align the optical axis of the star tracker with the payload optical reference alignment cube on the optical plate. To achieve this, two theodolites will be used. One of the theodolites will be referenced (by autocollimation) to the payload optical axis and the other will be referenced to the flat surfaces (by autocollimation, one surface at a time) on the star tracker. The reference angle between the reference cube and the star tracker surface would be measured by pointing the two theodolites at each other. Based on the angles the tip-tilt of the star tracker would be estimated and adjusted by shimming. The measurements and adjustments will be iterated until the required alignment is achieved. Post this alignment, the star tracker would be staked in its position. The final step in this phase is to remove the CGH and the CGH mount assembly from the optical plate.

\subsubsection{Phase 3: Detector alignment in vacuum}
\label{subsubsec:Phase_3}

\begin{figure}[htb]
    \centering
    \includegraphics[width=0.85\textwidth]{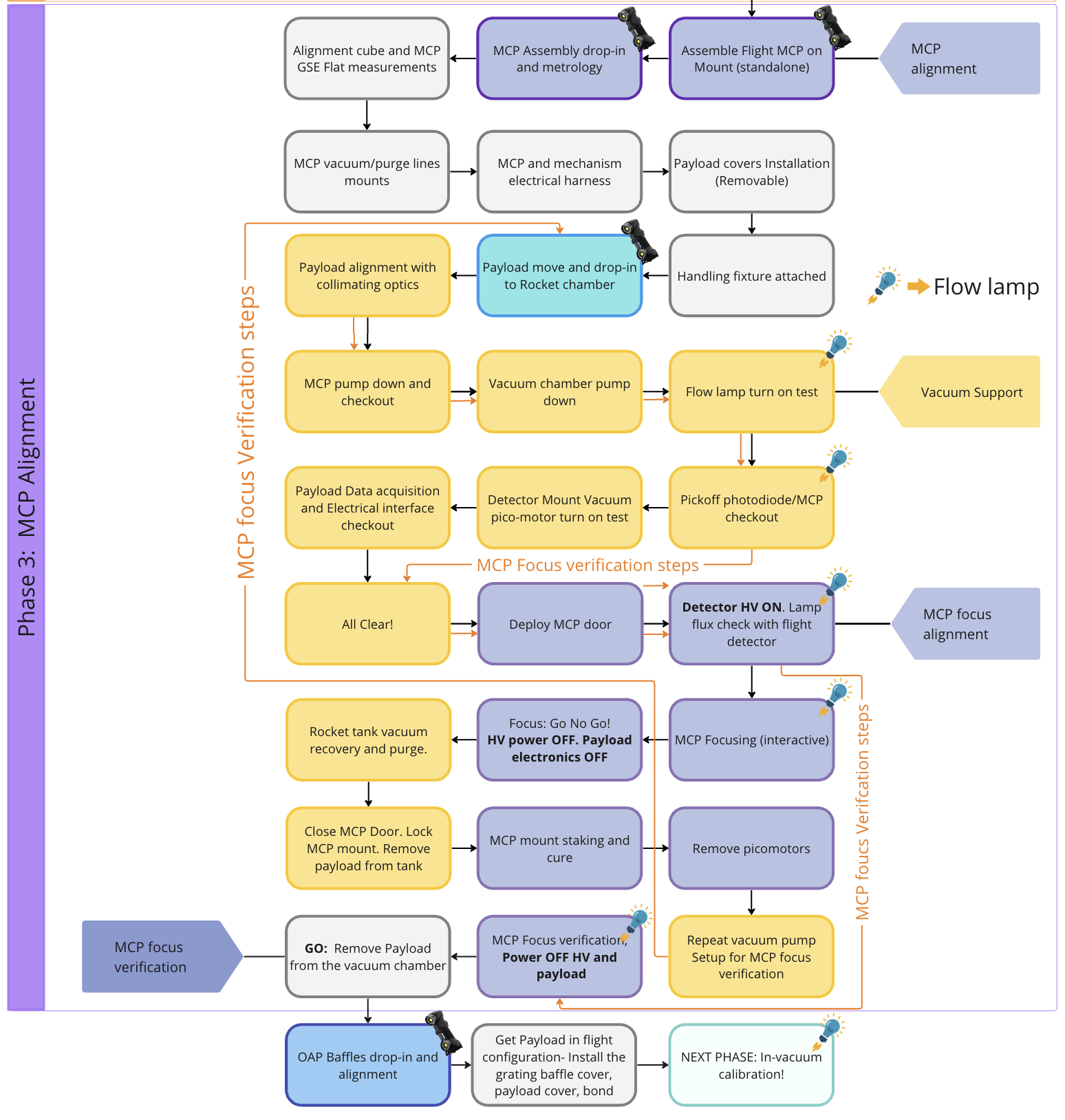}
    \caption{Phase 3 Flow Diagram: Step-by-step process for alignment of the detector with VUV collimator setup. Each block in the flow diagram represents an alignment activity with a unique procedure and activity ID. The color coding of the block is to group activities that are related to specific hardware or sequence of operations. The activities with the icon with Handyscan on the top right indicate that 3D scanner measurement will be done at this stage. The icon of a "bulb with rocket flames" indicates that the high voltage flow lamp was used for testing in that activity.}
    \label{fig:phase3}
\end{figure}

The alignment process flow for phase 3 is shown in Figure \ref{fig:phase3}.  In this phase, the MCP detector assembly is installed and positioned on the payload optical plate using coarse metrology with the 3D Scanner. During the assembly of the flight MCP housing, the location of the focal plane (MCP top surface) was measured to be within $\pm$ 50 $\mu$m  from the backplane of the MCP housing. During the installation of the detector, the backplane of the housing is used as a reference to get the coarse placement of the detector to within $\pm$ 100 $\mu$m. The precision alignment of the MCP can only be done in a vacuum with a VUV light source as the CsI Photocathode on the MCP is sensitive to humidity \cite{TREMSIN2000614}. The MCP assembly has a window that keeps the detector cavity closed and is only deployed if the detector is pumped down to vacuum or the instrument is under vacuum. For long-term storage, before assembly and during integration, the detector housing is back-filled with dry nitrogen and/or pumped down. 

\begin{figure}[htb]
    \centering
    \includegraphics[width=0.85\textwidth]{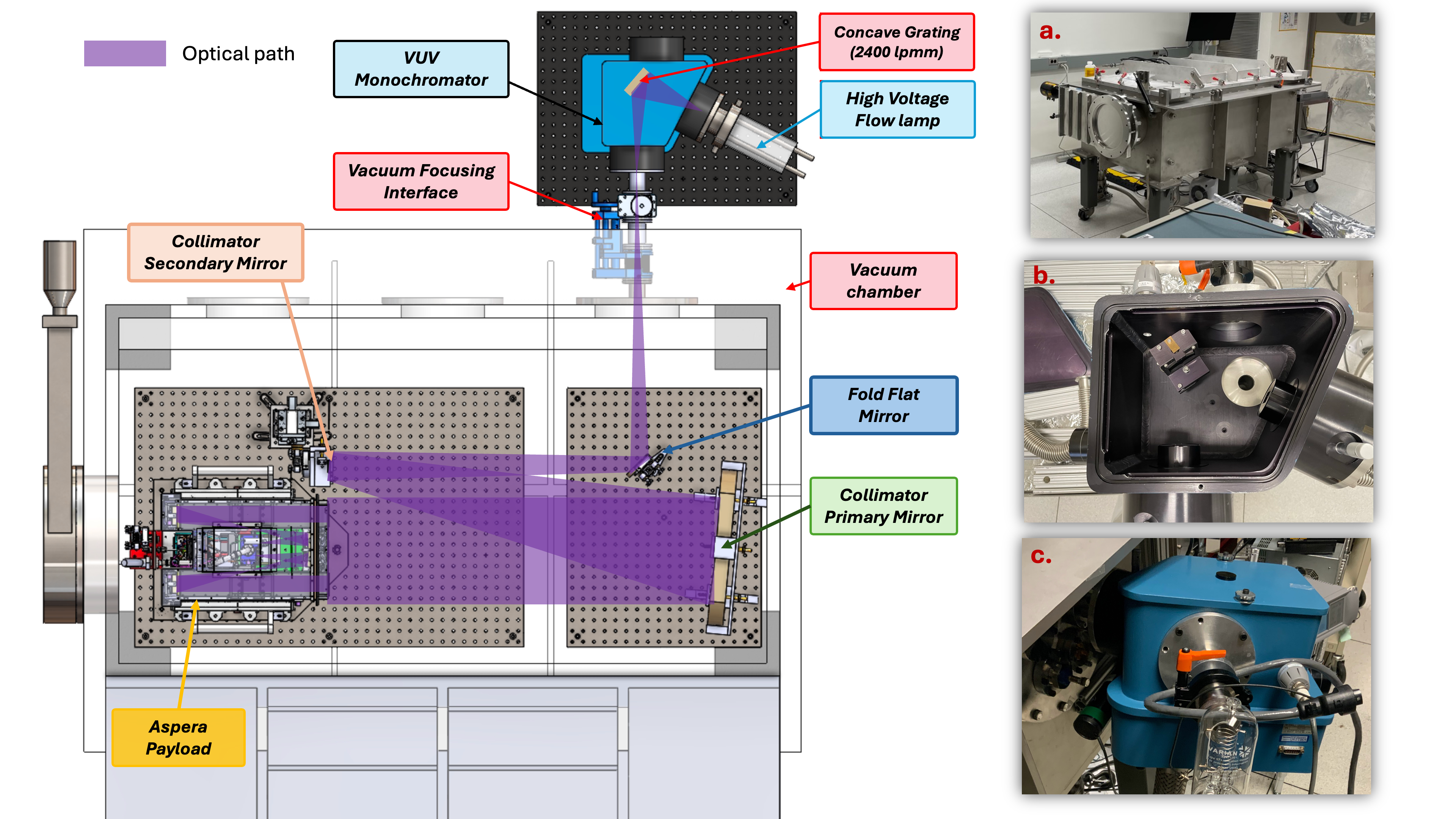}
    \caption{VUV test setup: (Left) Layout of the test setup for payload in vacuum for focusing the detector. The light source at the input of the Monochromator is a Windowless Hollow Cathode flow lamp that is fed with a mixture of Argon and Hydrogen gas. The monochromator exit slit is located at the focus of the collimator telescope. The collimated beam fills both apertures of \asp \ to allow simultaneous focusing of the two channels on the shared MCP detector. The panel on the right shows a) the vacuum calibration chamber used for the detector alignment, and b) \& c) the VUV monochromator that is being customized to produce monochromatic input for the collimating telescope.}
    \label{fig:vacuum_test_chamber}
\end{figure}

After the detector installation, the payload is placed in a vacuum test chamber for precision alignment of the detector. The test setup for the detector alignment is shown in Figure \ref{fig:vacuum_test_chamber}. The test setup consists of a two-mirror VUV collimator telescope, with a spherical primary and an aspheric secondary. The telescope optics are flight spares from the Hydrogen Polarimetric Explorer (HYPE) which is a far-ultraviolet polarimetric Spatial-Heterodyne spectrometer \cite{W_Harris_2010}. The original Al-MgF2 coating of these mirrors was stripped and replaced with a Platinum coating to improve throughput in the {\asp} bandpass. A flat mirror between the secondary and the focus of the telescope folds the beam to accommodate the setup in the vacuum chamber. At the input to the collimating telescope is a modified VM-502 VUV monochromator (from Acton Research Corporation) that is attached to the side of the vacuum chamber. The original 0.2 m focal length grating has been replaced with a platinum-coated concave grating with 2400 lines/mm groves that can resolve the emission lines from the High voltage (HV) flow lamp at the input slit of the monochromator. The flow lamp is operated at 2500 V with a mixture of Argon and Hydrogen to generate emission lines near \textit{Aspera’s} bandpass (Lyman $\beta$ at 1026 \AA \ and ArI at 1048 \AA). The exit slit of the monochromator is modified to mount a 200 $\mu$m pinhole aperture which corresponds to a $\sim$ 7-arcsecond source in the sky. The 14-inch wide collimated beam of the telescope can fill the two apertures of \asp \ simultaneously. 

Once the payload is placed in the vacuum tank and aligned to the collimator, the vacuum chamber is closed and pumped down. At this stage, the MCP window is deployed remotely to expose the detector. A High Voltage checkout is performed to ensure that the detector is working properly. The flow lamp is switched on with the monochromator centered at one of the two emission lines. The payload optics focus the monochromatic collimated beam on the flight detector. As the detector was only coarsely aligned, the collimated monochromatic input beam would create a defocused spot on the detector and the spot might be shifted from the designed position in the two channels due to detector tip-tilt. In addition, there may be some shift of spectral footprint (w.r.t. nominal) due to X and Y de-centers of the detector. The de-centers would be within the acceptable tolerance range after drop-in metrology as the footprint of the two channels on the detector is oversized with large margins.

 Three vacuum-compatible Picomotors\textsuperscript{\texttrademark} from Newport, attached to the flight mount with removable GSC, are used to move the detector in a vacuum. Each motor can move within $\pm$ 150 $\mu$m to adjust the defocus, tip, and tilt of the detector. The motors are controlled remotely and adjusted based on the feedback from the images taken with the MCP detector. The displacement for each motor is determined from the defocused image of the spots from the two channels at both wavelengths.. The modeled 80\% encircled energy (EE) for a combination of tip-tilt and defocus for the four spots will be compared against the measured 80\% EE of the spots in the images. To reduce the time in the process, a look-up-table (LUT) position and simulated images will be generated from the models. The adjustment will be stopped once the 80\% EE of the measured spots matches the simulated 80\% EE for all four spots. 

\section{Alignment simulation and Prototype alignment testing}
\label{sec:Prototype_testing}

\subsection{Simulation results from CGH alignment process}
\label{subsec:Simulation_CGH}
\begin{figure}[htb]
    \centering
    \includegraphics[width=0.85\textwidth]{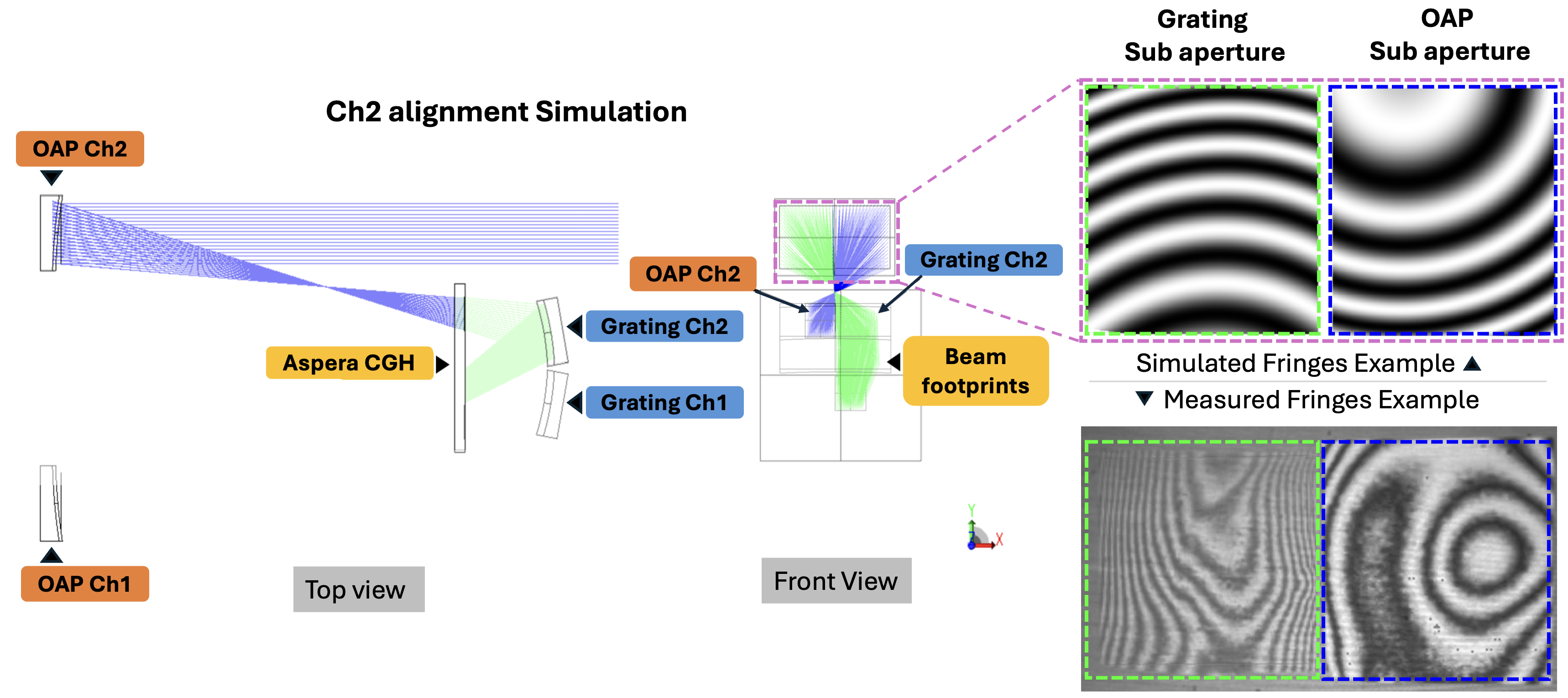}
    \caption{Simulation of CGH alignment. (Left) the optical path of the double pass Opticstudio model for \asp \ was used to simulate the alignment of a single channel. The footprint of the beam at the interferometer aperture and on the CGH is shown in the front view of the layout. (Top right) The OAP and grating were perturbed to a random starting position to simulate the fringes. (Bottom right) An example of measured fringes showing the fringes seen in the interferometer (fringe image has been rotated by 90 degrees to match with the layout of simulated fringes.) }
    \label{fig:cgh_simulation}
\end{figure}

The performance and sensitivity of the CGH alignment process were simulated using Ansys Zemax OpticStudio\textsuperscript{\textregistered}. The individual optics were moved to random positions within the design tolerance range using a Python interface with the Opticstudio model. The wavefront map for the interferometric measurements with CGH was simulated for each perturbed case for each optical element and Zernike polynomial coefficients derived by Zernike fitting the wavefront map on a predefined aperture. We also simulated the interferometer fringes for qualitative reference during the alignment process but these were not recorded for each case. Figure \ref{fig:cgh_simulation} shows the layout for the simulated alignment of Channel 2 (Ch2). A sample of simulated fringes is shown for a random position of OAP and Grating. For reference, an example of measured fringes from the prototype alignment test bench is shown. Based on these simulations, a wavefront error of $<
$300 nm RMS for the OAP return and $<$600 nm RMS for the grating return indicate that the optics are within the prescribed tolerance budget.

\subsection{Simulation results from detector focusing}
\label{subsec:Simulation_Detector_focusing}
The end-to-end model of the telescope and payload was used to simulate the quality of the collimated beam and its impact on the RMS spot size at the MCP detector of Aspera. The model is also used to estimate the sensitivity of the misalignment of the telescope optics and the alignment between the payload and the collimated beam. The collimation quality reduces with misalignment of the fold flat, primary, and secondary mirrors. While the misalignment between the payload and collimated beam of more than $\pm$ 15 arcseconds in the spatial direction would put the focused spot from the \asp \ OAPs outside the slit field of view. The f-number of the collimating telescope is $\sim$ 21.5. As this is a slow beam the effect of defocus on collimation quality is not significant. The spot sizes, for the two test wavelengths, on \asp \ detector do not change significantly even with a $\pm$ 5 mm shift in the position of the exit slit of the monochromator in defocus. 

\begin{figure}[htb]
    \centering
    \includegraphics[width=0.95\textwidth]{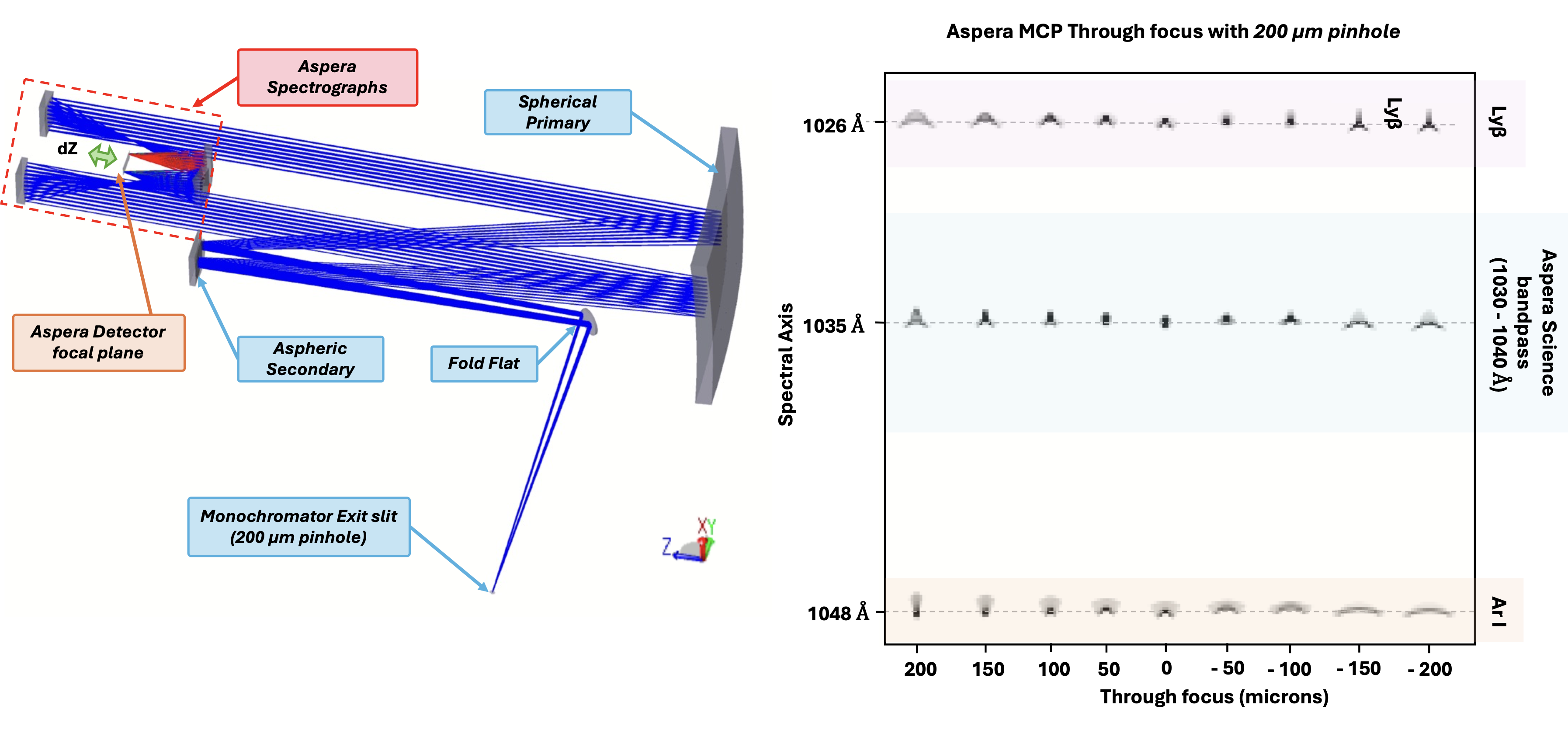}
    \caption{(Left) Simulation of detector alignment. The optical layout of the model used for simulating the detector alignment in vacuum). The collimated beam from the collimating telescope enters the aperture of the two channels for \asp. The collimated beam is dispersed and focused on the MCP detector by the two spectrograph channels. The monochromator at the entrance slit is tuned to one of the two test wavelengths. (Right) Simulated images through focus images of the 200 $\mu$m input pinhole at the \asp \ detector at the two test wavelengths are shown.}
    \label{fig:detector_simulation}
\end{figure}

Figure \ref{fig:detector_simulation} shows the optical layout of the simulation model and simulated through-focus images for one of the channels of \asp \ at three different wavelengths. The image simulation shows the on-axis spots for the Ly$\beta$ and Ar I lines for different defocused positions of the \asp \ MCP detector.  For fine alignment, the detector would be moved through focus and images would be acquired at different focus positions to determine the best focus location for the detector.

\subsection{Prototype alignment and testing}
\label{subsec:Prototype_testing}

The novel strategy of using in-situ 3D scanner meteorology and CGH posed considerable risks as these innovative methods have not been demonstrated in the context of a small-budget mission on a constrained timeline. Yet, the higher-risk posture had several advantages in terms of achieving the desired alignment tolerances at a low cost. To mitigate the risk of unknown problems that could arise out of using these methods, we developed a prototype model of the payload optical system. The prototype consists of replica optics with specifications similar to the flight optics. These optics were fabricated by diamond-turning the surface on aluminum substrates at Nanoform optics LLC. While these optics have the same dimensions and optical prescription; the surface roughness and figure error are limited by the capability of the diamond-turning process ($\sim$ 10 nm RMS microroughness, $\sim$ 90 nm RMS figure error). Additionally, the replica optics for the gratings have a smooth toroidal profile without the high-density grooves. The toroidal mirror can be used for our testing as the CGH nulls for grating are designed for the zeroth diffraction order. 

\begin{figure}[htb]
    \centering
    \includegraphics[width=0.85\textwidth]{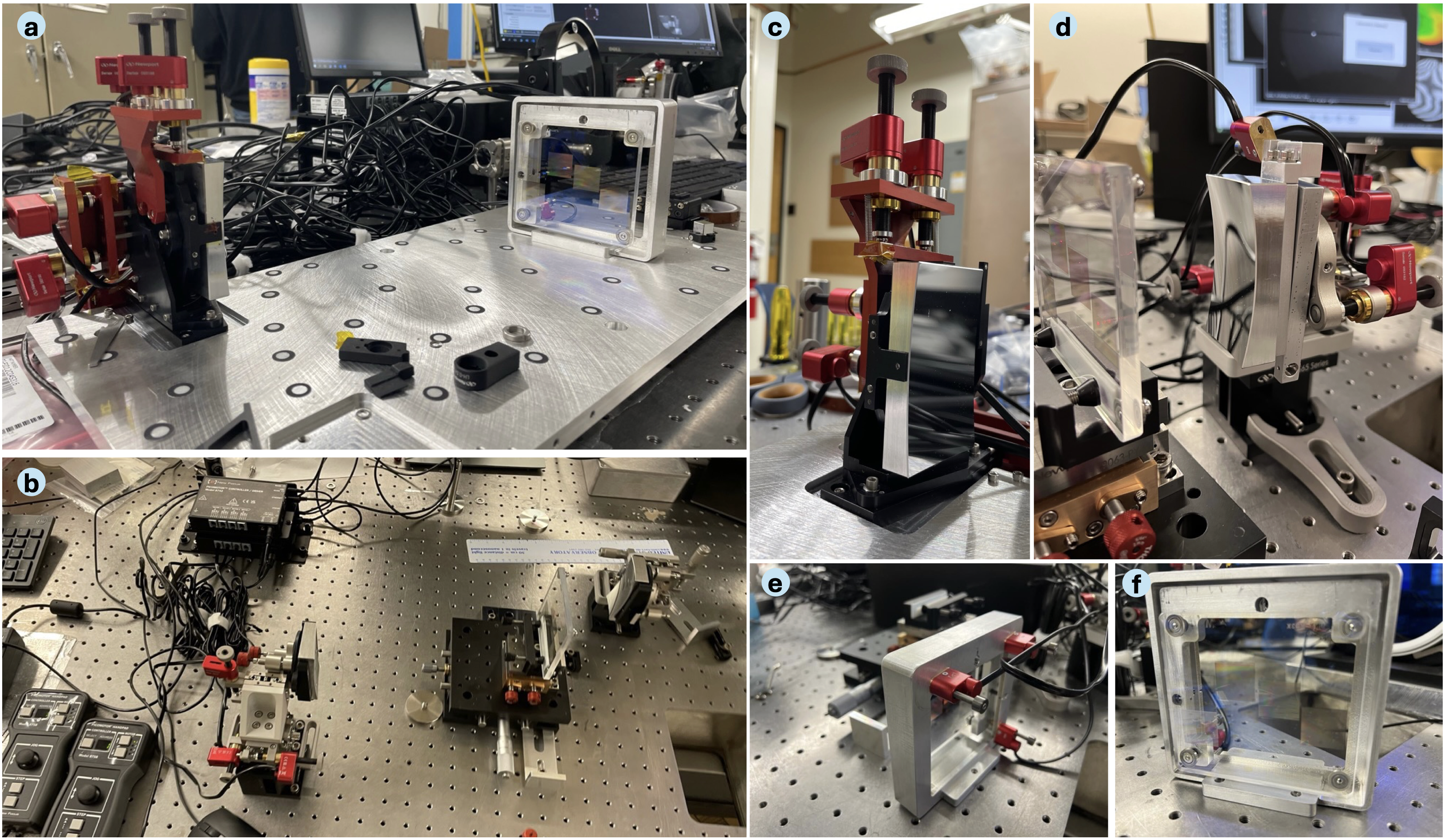}
    \caption{Evolution of the CGH alignment testbed and mounts. a) The COTS version of the alignment test bed with replica optics and CGH, b) the flight-like prototype test bed with the replica mounted on flight-light mounts with Picomotor GSC. c) Flight-like mount for OAP with Picomotor adjusters. d) Grating prototype mount made from COTS and custom parts. e,f) Prototype mount for the CGH with three Picomotor adjusters.}
    \label{fig:cgh_testbeds}
\end{figure}

We developed different versions of the prototype model as shown in Figure \ref{fig:cgh_testbeds}. For the prototype built in the early stages of payload design, we used a combination of commercial off-the-shelf (COTS) mounts and some custom parts to mount the optics for a single channel on an optical table. The CGH was mounted on a tip-tilt stage and aligned to the interferometer. After finalizing the design of flight hardware, the COTS mounts were replaced with flight-like mounts placed on a prototype of \asp's optical plate. In this configuration, the CGH assembly is mounted on a custom three-axis mount design to fit within the footprint of the system on the optical plate. The optical plate is mounted on the mass model version of the payload electronics bay by flight-like flexures. The flexure interface is torqued to reproduce the mounting stresses on the optical bench similar to the flight model. The progressive updates allowed us to test the alignment scheme as well as the assembly and performance of the mounts. 

\begin{figure}[htb]
    \centering
    \includegraphics[width=0.85\textwidth]{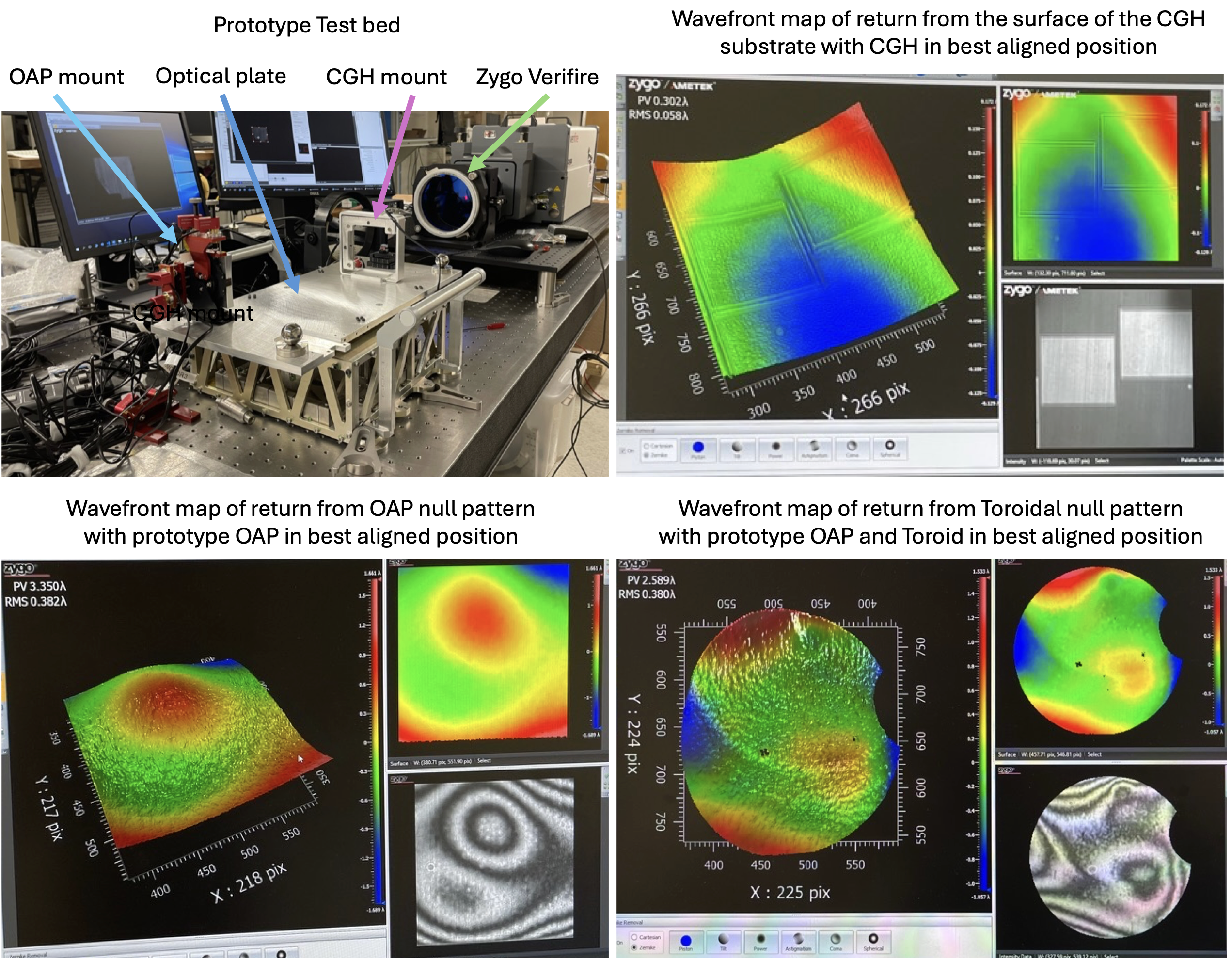}
    \caption{Prototype optics alignment with CGH. The prototype test bed with flight-like mounts and configuration (Top Left). Example of measured wavefront and fringes for the CGH (Top Right), the OAP (Bottom Left), and the gratings (Bottom Right). All three optics are well within the required alignment tolerances for the CGH alignment phase. The rms wavefront error for the CGH, OAP, and Gratings prototype wavefronts are 0.058 $lambda$, 0.382 $lambda$, and 0.380 $\lambda$ respectively which is within tolerance specification.}
    \label{fig:alignment_results}
\end{figure}

Figure \ref{fig:alignment_results} shows an example of wavefront measurements and fringes during prototype alignment activities. In this example, the CGH is aligned to within 30 nm RMS wavefront error most of which comes from residual tilts. The wavefront error for the OAP and the grating sub-apertures is within 240 nm. Most of the wavefront error is contributed by the figure of the aluminum prototype mirrors as the RMS wavefront error does not decrease any further with rigid body movements in any direction.

\section{Future work and Summary}
\label{sec:status_futurework}
\subsection{Flight alignment status}
At the time of writing, the \asp \ flight model is being integrated at the Lunar Planetary Lab cleanroom facility at the University of Arizona. All the optics have been fabricated and delivered. The optical coatings for the OAPs are done, and the coating for the gratings are in progress. Most of the optomechanical components for flight and GSC have been fabricated. The same CGH used in prototype testing would be used for flight alignment with an updated and precision-cleaned mount. The optical alignment for the flight model is expected to begin in early July 2024 and finish by the end of 2024. The payload calibration and environmental testing will completed in early 2025 and the payload will be delivered to SFL for integration into the spacecraft. 

\subsection{Summary}
\asp \ is a SmallSat mission in the FUV designed to observe the {\OVI} emission from the CGM of nearby galaxies. The small form factor and contamination sensitivity of the spectrograph pose several challenges in the alignment and performance verification of the spectrograph optics. This paper describes the strategies and processes developed for the end-to-end optical alignment of the spectrograph. The alignment is done in a multi-phase, multi-step process with each phase beginning with the coarse alignment using an industrial meteorology-grade 3D scanner. The high-speed acquisition and high-accuracy direct scan to CAD measurements allow for precision contactless meteorology suitable for contamination-sensitive VUV optics. Placing the optomechanical assembly with high precision reduces the complexity of the mount by reducing the travel range required during fine alignment. The optical alignment is done using a novel CGH designed to meet the tight alignment tolerances. The flight detector is aligned to the optics aligned by the CGH in a VUV test setup. The performance verification after the alignment and staking of the optical components will be done using the same VUV test setup to ensure that the payload is tested in a flight-like configuration. This work combines multiple innovative optomechanical and optical alignment methods to simplify the challenging alignment of VUV spectrographs. 

\acknowledgments 

This research is supported by the NASA Astrophysics Pioneers Program, Aspera Revealing the Diffuse Universe (80NSSC22M0081). The authors would also like to acknowledge partial funding support from the NASA Future Investigators in NASA Earth and Space Science and Technology (FINESST) Grant (80NSSC21K2050). Team \asp \ would like to acknowledge the support from the NASA Astrophysics Division Program Office, the University of Arizona Space Institute (UASI), the Steward Observatory, the Lunar Planetary Laboratory, Zygo Corporation, and Arizona Optical Metrology LLC (AOM). We would also like to thank Dr. Jared Males, Dr. Laird Close, and their research group at Steward Observatory for sharing their valuable resources.  


\bibliography{main} 

\begin{thebibliography}{10}

\bibitem{Tumlinson_2017}
Tumlinson, J., Peeples, M.~S., and Werk, J.~K., ``{The Circumgalactic Medium},'' {\em Annual Review of Astronomy and Astrophysics}~{\bf 55},  389–432 (Aug. 2017).

\bibitem{Keres_2005}
{Kere{\v{s}}}, D., {Katz}, N., {Weinberg}, D.~H., and {Dav{\'e}}, R., ``{How do galaxies get their gas?},'' {\em Monthly Notices of the Royal Astronomical Society}~{\bf 363},  2--28 (Oct. 2005).

\bibitem{Fielding_2017}
Fielding, D., Quataert, E., McCourt, M., and Thompson, T., ``{The impact of star formation feedback on the circumgalactic medium},'' {\em Monthly Notices of the Royal Astronomical Society}~{\bf 466},  3810--3826 (May 2017).
\newblock Publisher Copyright: {\textcopyright} 2016 The Authors Published by Oxford University Press on behalf of the Royal Astronomical Society.

\bibitem{Putman2017}
Putman, M.~E.,  [{\em {An Introduction to Gas Accretion onto Galaxies}}{\nolinebreak\hspace{0.1em}]},  1--13, Springer International Publishing, Cham (2017).

\bibitem{Chung2021}
Chung, H., Vargas, C.~J., Hamden, E.~T., McMahon, T., Gonzales, K.~L., Khan, A.~R., Agarwal, S., Bailey, H., Behroozi, P., Brendel, T., Choi, H., Connors, T., Corlies, L., Corliss, J., Dettmar, R.-J., Dolana, D., Douglas, E.~S., Guzman, J., Hamara, D., Harris, W., Harshman, K., Hergenrother, C., Hoadley, K., Kidd, J., Kim, D.~W., Li, J.~S., Montoya, M., Sauve, C., Schiminovich, D., Selznick, S., Siegmund, O., Ward, M., Wolcott, E.~M., and Zaritsky, D., ``{Aspera: the UV SmallSat telescope to detect and map the warm-hot gas phase in nearby galaxy halos},'' (Aug. 2023).

\bibitem{Feldman_2001}
Feldman, P.~D., Sahnow, D.~J., Kruk, J.~W., Murphy, E.~M., and Moos, H.~W., ``{High-resolution FUV spectroscopy of the terrestrial day airglow with the Far Ultraviolet Spectroscopic Explorer},'' {\em Journal of Geophysical Research: Space Physics}~{\bf 106}(A5),  8119--8129 (2001).

\bibitem{Otte_2006}
Otte, B. and Van Dyke~Dixon, W., ``{The Far Ultraviolet Spectroscopic ExplorerSurvey of OVI Emission in the Milky Way},'' {\em The Astrophysical Journal}~{\bf 647},  312–327 (Aug. 2006).

\bibitem{Agarwal+24}
{Agarwal}, S. and {et al. }, ``{Straylight analysis and baffle design for the Aspera UV SmallSat astrophysics mission},'' in [{\em Space Telescopes and Instrumentation 2024: Ultraviolet to Gamma Ray}{\nolinebreak\hspace{0.1em}]},  {\em Proc. of SPIE} {\bf 13093 (in Prep)} (2024).
\newblock “Manuscript in preparation”.

\bibitem{Chung+24}
{Chung}, H. and {et al. }, ``{Payload design and development of Aspera: the UV SmallSat mission for mapping warm-hot phase gas in nearby galaxy halos},'' in [{\em Space Telescopes and Instrumentation 2024: Ultraviolet to Gamma Ray}{\nolinebreak\hspace{0.1em}]},  {den Herder}, J.-W.~A., {Nikzad}, S., and {Nakazawa}, K., eds., {\em Proc. of SPIE} {\bf 13093 (in Prep)} (2024).

\bibitem{choi2023}
Choi, H., Chung, H., Vargas, C., Khan, A.~R., Corliss, J., and Kim, D., ``{Tolerance and alignment study of two-channel telescope spectroscopy for Aspera},'' in [{\em Astronomical Optics: Design, Manufacture, and Test of Space and Ground Systems IV}{\nolinebreak\hspace{0.1em}]},  Hull, T.~B., Kim, D., and Hallibert, P., eds.,  {\bf PC12677},  PC126770D, International Society for Optics and Photonics, SPIE (2023).

\bibitem{Plesseria2001}
{Plesseria}, J.~Y., {Henrist}, M., and {Doyle}, D., ``{A New Guideline on Contamination Control for Space Optical Instruments},'' in [{\em Fourth International Symposium Environmental Testing for Space Programmes}{\nolinebreak\hspace{0.1em}]},  {Sch{\"u}rmann}, B., ed., {\em ESA Special Publication} {\bf 467},  129 (Jan. 2001).

\bibitem{Tribble1996}
Tribble, A.~C., Boyadjian, B., Davis, J., Haffner, J., and McCullough, E., ``{Contamination control engineering design guidelines for the aerospace community},'' in [{\em Optical System Contamination V, and Stray Light and System Optimization}{\nolinebreak\hspace{0.1em}]},  Breault, R.~P., Pompea, S.~M., Glassford, A. P.~M., Breault, R.~P., and Pompea, S.~M., eds.,  {\bf 2864},  4 -- 15, International Society for Optics and Photonics, SPIE (1996).

\bibitem{Melso2024}
{Melso}, N., {Khan}, A., {Tanquary}, H., {Garcia}, E., {Truong}, D., {Yescas}, N., {Park}, S., {Uppnor}, S., {Vargas}, C., {Chung}, H., {Mcmahon}, T., {Hoadley}, K., {Hamden}, E., {Corliss}, J., {Verts}, B., {Agarwal}, S., {Augustin}, R., {Behroozi}, P., {Bradley}, H., {Brendel}, T., {Burchett}, J., {Martinez Catillo}, J., {Chambers}, J., {Choi}, H., {Corliss}, L., {Coronado}, F., {Davis}, G., Ralf-J{\"u}rgen, D., {Douglas}, E., {Ghidoli}, G., {Goodwin}, A., {Hamara}, D., {Harris}, W., {Hergenrother}, C., {Howk}, C., {Keppler}, M., {Kidd}, J., {Kim}, D.~W., {Li}, J., {Lochhaas}, C., {Noenickx}, J., {Noriega}, G., {Pencha}, R., {Sauve}, C., {Schiminovich}, D., {Selznick}, S., {Oswald}, S., {Su}, R., {Wolcott}, E., and {Zaritsky}, D., ``{Contamination Control for the Aspera FUV SmallSat},'' in [{\em Ultraviolet to Gamma Ray}{\nolinebreak\hspace{0.1em}]},  {\em Astronomical Telescopes and Instrumentation (SPIE) Conference Series} (2024).
\newblock “Manuscript in preparation”.

\bibitem{Wilbrandt_2014}
Wilbrandt, S., Stenzel, O., Nakamura, H., Wulff-Molder, D., Duparré, A., and Kaiser, N., ``{Protected and enhanced aluminum mirrors for the VUV},'' {\em Applied optics (2004)}~{\bf 53}(4),  A125–A130 (2014).
\newblock ObjectType-Article-1.

\bibitem{Fleming:17}
Fleming, B., Quijada, M., Hennessy, J., Egan, A., Hoyo, J.~D., Hicks, B.~A., Wiley, J., Kruczek, N., Erickson, N., and France, K., ``{Advanced environmentally resistant lithium fluoride mirror coatings for the next generation of broadband space observatories},'' {\em Appl. Opt.}~{\bf 56},  9941--9950 (Dec 2017).

\bibitem{TREMSIN2000614}
Tremsin, A., Ruvimov, S., and Siegmund, O., ``{Structural transformation of CsI thin film photocathodes under exposure to air and UV irradiation},'' {\em Nuclear Instruments and Methods in Physics Research Section A: Accelerators, Spectrometers, Detectors and Associated Equipment}~{\bf 447}(3),  614--618 (2000).

\bibitem{khan2024calibration}
Khan, A.~R., Chung, H., Vargas, C.~J., Hamden, E.~T., McMahon, T., Tanquary, H., Coronado, F., Melso, N., Hamara, D., Keppler, M., Li, J.~S., Corliss, J.~B., Agarwal, S., Choi, H., Garcia, E., Ghidoli, G., Hergenrother, C., Kidd, J., Kim, D., Noenickx, J., Park, S., Selznick, S., Truong, D., Uppnor, S., Verts, B., Yescas, N., Batkis, M., Harris, W.~M., Hennessy, J.~J., Hoadley, K., de~Marcos, L. V.~R., Martin, A., Quijada, M.~A., and Siegmund, O. H.~W., ``{An overview of the on-ground and in-orbit calibration of Far-Ultraviolet Spectrographs for the Aspera SmallSat},'' in [{\em SPIE Astronomical Telescopes + Instrumentation}{\nolinebreak\hspace{0.1em}]},  (June 2024).
\newblock “Paper 13093-115, Manuscript in preparation”.

\bibitem{creaform_handyscan_2024}
{Creaform}, ``Handyscan 3d technical specifications,'' (2024).
\newblock Accessed: 2024-05-29.

\bibitem{Khan_2025}
Khan, A. et~al., ``{Optical alignment and performance verification of Aspera Spectrographs},'' Unpublished manuscript, in Preparation (expected 2025).

\bibitem{W_Harris_2010}
B{\'e}tr{\'e}mieux, Y., Corliss, J., Vincent, M.~B., Vincent, F.~E., Roesler, F.~L., and Harris, W.~M., ``{Description and ray-tracing simulations of HYPE: a far-ultraviolet polarimetric spatial-heterodyne spectrometer},'' in [{\em Space Telescopes and Instrumentation 2010: Ultraviolet to Gamma Ray}{\nolinebreak\hspace{0.1em}]},  Arnaud, M., Murray, S.~S., and Takahashi, T., eds.,  {\bf 7732},  77322C, International Society for Optics and Photonics, SPIE (2010).

\end{thebibliography}
\bibliographystyle{spiebib} 

\end{document}